\documentclass[twocolumn, revtex4, numberedappendix,iop]{openjournal}

\usepackage{amsmath, amsfonts, amssymb} 
\usepackage{graphicx} 
\usepackage[colorlinks, linkcolor=blue, citecolor=blue]{hyperref} 
\usepackage{xspace} 
\usepackage{xcolor} 
\usepackage{upgreek} 
\usepackage{tabularx} 

\newcommand{\dotp}[2]{\mathbf{#1}\cdot\mathbf{#2}}

\newcommand{\ft}[3]{\int{#1}e^{-{\rm i}\dotp{#2}{#3}}\,\mathrm{d}^3{#3}}

\newcommand{\average}[1]{\langle#1\rangle}

\newcommand{\sform}[2]{{#1}\times10^{#2}}
\renewcommand{\pi}{\uppi}

\newcommand{\order}[1]{\mathcal{O}(#1)}

\newcommand{\diff}{\mathrm{d}}

\newcommand{\ba}{\begin{eqnarray}}
\newcommand{\ea}{\end{eqnarray}}
\newcommand{\be}{\begin{equation}}  
\newcommand{\ee}{\end{equation}}

\newcommand{\Mpc}{\,h^{-1}\mathrm{Mpc}}

\newcommand{\iMpc}{\,h\mathrm{Mpc}^{-1}}

\newcommand{\Msun}{\,h^{-1}\mathrm{M_\odot}}

\newcommand{\cm}{\,\mathrm{cm}}

\newcommand{\Om}{\Omega_\mathrm{m}}

\newcommand{\Bnl}{\beta^\mathrm{nl}}

\newcommand{\eg}{e.g.,\xspace}
\newcommand{\ie}{i.e.\xspace}

\newcommand{\nbody}{$N$-body\xspace}
\newcommand{\LCDM}{$\Lambda$CDM\xspace}
\newcommand{\halofit}{\textsc{halofit}\xspace}
\newcommand{\DQ}{\textsc{dark quest}\xspace}
\newcommand{\wmap}{\textit{WMAP}\xspace}
\newcommand{\hmcode}{\textsc{hmcode}\xspace}

\newcommand{\halomod}{\textsc{halomod}\xspace}
\newcommand{\darkemu}{\textsc{darkemu}\xspace}
\newcommand{\camb}{\textsc{camb}\xspace}

\newcommand{\react}{\textsc{react}\xspace}
\newcommand{\python}{\textsc{python}\xspace}

\newcommand{\link}[1]{\href{#1}{#1}}

\newcommand{\pyhalomodel}{\href{https://github.com/alexander-mead/pyhalomodel}{\sc pyhalomodel}}

\definecolor{purple}{RGB}{156,81,182}

\definecolor{applegreen}{rgb}{0.55, 0.71, 0.0}

\newcommand{\citecaption}[1]{\protect\cite{#1}} 

\def\mnras{MNRAS}
\def\apj{ApJ}
\def\apjl{ApJ Let.}
\def\prd{Phys. Rev. D}
\def\aap{A\&A}
\def\jcap{J. Cosmol. Astropart. Phys.}
\def\apjs{ApJS}

\begin{document} 

\title{The halo model for cosmology: A pedagogical review} 

\author{M. Asgari$^{1,2,3}$\footnote{m.asgari@hull.ac.uk}, A. J. Mead$^{4,3}$\footnote{alexander.j.mead@googlemail.com}, and C. Heymans$^{3,4}$}
\affiliation{
$^1$E. A. Milne Centre, University of Hull, Cottingham Road, Hull, HU6 7RX, UK\\
$^2$Centre of Excellence for Data Science, AI, and Modelling (DAIM), University of Hull, Cottingham Road, Hull, HU6 7RX, UK\\
$^3$Institute for Astronomy, University of Edinburgh, Royal Observatory, Blackford Hill, Edinburgh, EH9 3HJ, UK\\
$^4$Ruhr University Bochum, Faculty of Physics and Astronomy, Astronomical Institute (AIRUB), German Centre for Cosmological Lensing, 44780 Bochum, Germany\\
}

\begin{abstract}
We present a pedagogical review of the halo model, a flexible framework that can describe the distribution of matter and its tracers on non-linear scales for both conventional and exotic cosmological models.  We start with the premise that the complex structure of the cosmic web can be described by the sum of its individual components: dark matter, gas, and galaxies, all distributed within spherical haloes with a range of masses.  The halo properties are specified through a series of simulation-calibrated ingredients including the halo mass function, non-linear halo bias and a dark matter density profile that can additionally account for the impact of baryon feedback.  By incorporating a model of the galaxy halo occupation distribution, the properties of central and satellite galaxies, their non-linear bias and intrinsic alignment can be predicted.  Through analytical calculations of spherical collapse in exotic cosmologies, the halo model also provides predictions for non-linear clustering in beyond-$\Lambda$CDM models.  The halo model has been widely used to model observations of a variety of large-scale structure probes, most notably as the primary technique to model the underlying non-linear matter power spectrum.  By documenting these varied and often distinct use cases, we seek to further coherent halo model analyses of future multi-tracer observables.  This review is accompanied by the release of \pyhalomodel, flexible software to conduct a wide range of halo-model calculations.
\end{abstract}

\keywords{Cosmology,  large-scale structures}

\maketitle 
\tableofcontents

\section{Introduction}
\label{sec:introduction}

On large scales, and at early times, matter fluctuations are small and can be described using linear perturbation theory; the evolution of small perturbations can be solved analytically.  Once fluctuations become more developed, however, their properties can no longer be explained by linearised equations, and instead a full non-linear treatment is needed.  The halo model provides an intuitive way to approximate the matter distribution in the non-linear regime.  It posits that all matter resides in haloes, which are $\mathcal{O}(100)$ times denser than the cosmological average -- a view that has been largely corroborated by numerical simulations.  Once the properties and distribution of these haloes are known, one can estimate the statistical properties of the matter distribution in the cosmos.  To be concrete,  power spectra for matter and its tracers can be understood as the sum of two components: inter-halo (two-halo) and intra-halo (one-halo) clustering.  The halo model, therefore, can (and has) been used in analysing cosmological data from various probes of large-scale structure.  

The statistical properties of any tracer of matter can also be modelled,  provided that the connection between the tracer and host haloes is known.  If haloes are taken to be the sites of galaxy formation, all that is needed to model the galaxy clustering signal is how galaxies occupy haloes of different masses. The problem then can be split into how galaxies cluster within the same halo and how different haloes, which might include varying numbers of galaxies, cluster with respect to each other. In principle, the same logic can be applied to \emph{any} tracer of the large-scale structures,  as long as the signal from the tracer emanates from haloes.  For example,  the thermal Sunyaev-Zel'dovich (tSZ) effect is sourced by electron pressure,  which is at its most intense within haloes, and so reasonable models for tSZ clustering, and its cross-correlation with other tracers, may be derived using the halo model.  

The halo properties that are required to make a prediction using the halo model are the halo bias (how haloes cluster relative to matter), halo mass function (number density of haloes with different masses), and halo profile (how matter or its tracers are distributed within a halo). These ingredients are most often extracted from numerical simulations and (sometimes) calibrated across a range of cosmological parameters.  It is usual to assume that haloes are linearly biased,  spherical objects with properties that are only a function of the halo mass,  although these restrictions can be relaxed.  We call this method of using the halo model the ``analytical approach".

There is a second approach to using the halo model, the ``simulation-based approach":  Here, haloes are identified in a simulation and then `painted' with a specific tracer (\eg galaxies), such that the desired clustering properties can be directly measured.  While the analytical approach is quicker and more flexible, the simulation-based approach is potentially more accurate,  but it is slower and requires \nbody simulations.  This can become a problem in cosmological analyses where a wide range of  parameters and/or cosmological models need to be explored.  Once the analytical approach is adjusted to reach a desired accuracy,  it can be extended more readily to other cosmological parameters and/or models compared to the simulation based approach.

The halo model has been used in one form or another to analyse data from weak gravitational lensing by large-scale structure,  known as cosmic shear.  This is because cosmic shear relies on information from non-linear matter distribution.  \halofit \citep{Smith2003,Takahashi2012} and \hmcode \citep{Mead2015a,Mead2021a} which have their roots in the halo-model approach (see section~\ref{sec:modelling_matter}) form the basis of all primary analysis of recent cosmic shear data: Canada France Hawaii Telescope lensing Survey \citep[CFHTLenS,][]{Heymans2013,Joudaki2017},  Deep Lens Survey \citep[DLS,][]{Jee2013},  Kilo Degree Survey  \citep[KiDS,][]{Hildebrandt2017, Hildebrandt2020, Asgari2021},  Dark Energy Survey \citep[DES,][]{Troxel2018,Amon2022,Secco2022} and Hyper Suprime-Cam \citep[HSC,][]{Hikage2019,Hamana2020}.

Data from galaxy clustering and the cross-correlation between weak lensing and galaxy clustering,  known as galaxy--galaxy lensing,  have been analysed with a flexible halo model approach to capture information from smaller scales \citep[for example][]{Cacciato2013, More2015, Miyatake2022b, Dvornik2023}.  \cite{Troester2022} applied the halo model formalism of \cite{Mead2020} to the cross-correlation between tSZ and weak lensing in a cosmological analysis.  The halo model can also predict the intrinsic alignments of galaxies,  which is a prominent astrophysical systematic in cosmic-shear studies \citep{Schneider2010,Fortuna2021}. 

While data from modern \emph{galaxy} surveys is most often modelled using a halo model applied to, or directly calibrated against simulations,  most analyses concerning cross-correlations of different tracers use the 'vanilla' halo model as a first modelling effort \citep[\eg][]{Komatsu2002, Hill2014, Hurier2014, Battaglia2015, Troester2017, Feng2017, Osato2018, Wolz2019, Koukoufilippas2020, Yan2021}.   The limitations of the analytical halo-model approach are appreciated and accounted for when it comes to modelling the galaxy and matter cross-correlation (galaxy--galaxy lensing), but less well appreciated for other cross correlations where the modelling and data are less mature.  As we will discuss later, these limitations are a strong function of exactly what one is trying to model, particularly of the relationship between tracer strength and halo mass. For these reasons, we feel that a pedagogical review is useful and timely.

Extracting information about cosmological parameters and galaxy-occupation statistics via the halo model has been historically challenging. The simplicity of the analytical halo model, which is appealing from the perspective of understanding, can become a problem as the data to which it is exposed increases in quality. The halo model should be calibrated against simulations to check its accuracy, and a failure to do so may result in incorrect parameter constraints where real signal is mistaken for some modelling deficiency. There have been many attempts to use the halo model to constrain cosmological and galaxy--halo parameters \citep[\eg][]{Tinker2005, vandenBosch2013, More2013, Cacciato2013, More2015a, Leauthaud2017,Zacharegkas2022}. Recent attempts to include non-linear halo bias \citep[][]{Nishimichi2019, Mead2021b} within the halo model provide a promising way to improve the accuracy of the halo model to the extent that it can be trusted to recover cosmological parameters and information about the galaxy--halo connection from current \citep[\eg][]{Mahony2022,Miyatake2022b, Dvornik2023} and forthcoming survey data.  Another powerful property of the halo model is that its ingredients can be directly calibrated against real observations,  by allowing for parameters to vary freely in likelihood analyses \citep[see for example][]{Gu2023}. 

The halo model first came to cosmological prominence at the beginning of the millennium \citep{Seljak2000, Ma2000b, Peacock2000}  following earlier work by \cite{Scherrer1991} and, after a flurry of related publications, was first reviewed by \cite{Cooray2002}. As far as we know, there has been no subsequent attempt to review the halo model. Here, we attempt to provide a pedagogical summary of the modern uses of the halo model with a focus on its cosmological applications and the modelling of the auto and cross power spectra of different fields.

The structure of this review is as follows: In Section~\ref{sec:basics} we provide a comprehensive derivation of the halo-model power spectrum in order to highlight the assumptions lurking behind the model. In Section~\ref{sec:ingredients} we discuss the ingredients that are necessary to make the halo model predictive in the case of the matter distribution and how it is modelled in practice, while in Section~\ref{sec:tracers} we see how the halo model can be extended for tracers of matter, such as galaxies. In Section~\ref{sec:non_standard} we discuss some non-standard approaches and improvements to halo modelling that have appeared over the years. In Section~\ref{sec:altcosmo} we discuss applications to cosmologies beyond the standard \LCDM. In Section~\ref{sec:software} we detail the publicly-available software for performing halo-model calculations, and finally we summarise in Section~\ref{sec:summary}.

\section{The halo model basics}
\label{sec:basics}

\begin{figure*}
\begin{center}
\includegraphics{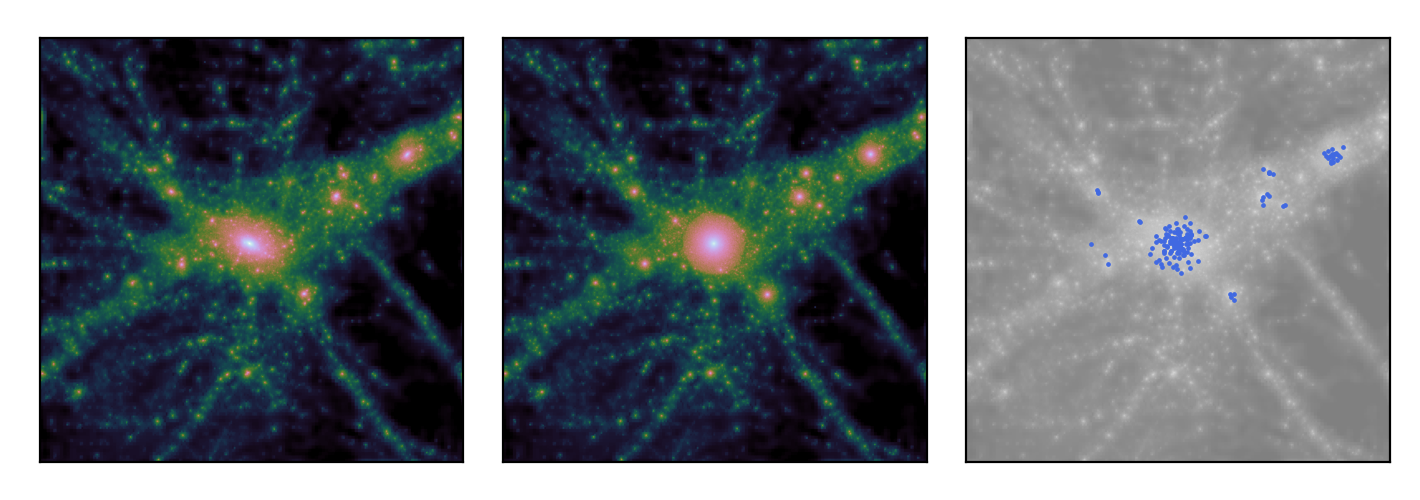}
\end{center}
\caption{A schematic visualisation of the halo-model process. The left-hand panel shows the matter density field in a $25\times25\times5\,(h^{-1}\mathrm{Mpc})^3$ region of an \nbody simulation, centred on a massive ($\sim10^{14.5}\Msun$) halo identified at $z=0$. The central panel shows the result of isolating all haloes identified in the simulation and replacing these with idealised spherical haloes of the same mass. The right-hand panel shows the result of populating these haloes with galaxies according to a simple galaxy-occupation prescription.}
\label{fig:visualisation}
\end{figure*}

At the core of the halo model is the approximation that we can fully describe the complex structure of the cosmic web simply as a sum of its individual components: dark matter, gas, and galaxies, all distributed in haloes. The complexity then translates into the problem of how to model the profile, mass distribution and bias of these haloes relative to the underlying (linear-theory) matter distribution, and how to accurately populate haloes with gas and galaxies. This `mapping information' can be motivated by, or extracted from, numerical simulations, leading to a flexible analytical model of the statistical properties of the cosmic web. By confronting this model with observations, the halo-model approach can provide constraints on the underlying cosmology of the Universe, in addition to providing unique insight into the halo--galaxy connection, and furthering our understanding of galaxy formation and evolution. A schematic of the halo-model approach is shown in Fig.~\ref{fig:visualisation}, where the first panel shows the density field in an \nbody simulation, the second panel shows this replaced by a spherical halo approximation, and the final panel shows a possible galaxy distribution.

\onecolumngrid
\hrulefill

\subsection{Standard derivation}
\label{sec:derivation}

We start by defining a field in real space, $\theta_{\rm u}(\bold{x})$, where  $\bold{x}$ is the three-dimensional comoving position and the label ${\rm u}$ stands for the field we are interested in modelling. Fields are also a function of time, usually parameterised via $z$, but we suppress this argument here and throughout this paper to make the notation less cluttered. Examples of such fields would be `matter', `halo' or `galaxy' over-densities that vary from place-to-place in the Universe.  
We make the assumption that everything in our field is contained within haloes distributed throughout the space with a spherically symmetric profile, $W_{{\rm u},i}$, centred at position $\bold{x}_i$, such that
\begin{equation}
\theta_{\rm u}(\mathbf{x})=\sum_i N_i\;W_{{\rm u},i}(|\mathbf{x}-\mathbf{x}_i|)\;,
\label{eq:theta_haloes}
\end{equation}
where the sum runs over all volume elements and $N_i=\lbrace0,1\rbrace$ determines whether there is a halo centre in that volume element.  
The Fourier transform of the field, in terms of comoving wavenumber $\bold{k}$, is given by
\begin{equation}
\hat{\theta}_{\rm u}(\bold{k})=\ft{\theta_{\rm u}(\bold{x})}{\bold{k}}{\bold{x}}\ ,
\end{equation}
With the variable change $\mathbf{r}=\bold{x}-\bold{x}_i$,
\begin{align}
\label{eq:ffield}
\hat{\theta}_{\rm u}(\bold{k})=\sum_i\mathrm{e}^{-{\rm i}\dotp{k}{x}_i}\; N_i \int  W_{{\rm u},i}(|\bold{r}|) \mathrm{e}^{-{\rm i} |\bold{k}||\bold{r}|\cos\theta} d^3 \bold{r} 
= \sum_i\mathrm{e}^{-{\rm i}\dotp{k}{x}_i} N_i\; \hat{W}_{\mathrm{u},i}(k)\ ,
\end{align}
where we have recognised the integral as the Fourier transform of the halo profile, $\hat{W}_{{\rm u},i}(k)$.  The spherical symmetry of the halo profile allows us to integrate over the angular co-ordinates such that the Fourier transform of the halo profile is given by
\begin{equation}
\hat{W}_{{\rm u},i}(k)=\int_0^\infty\frac{\sin(kr)}{kr}\;W_{{\rm u},i}(r)\;4\pi r^2\,\diff r\ .
\label{eq:window_function}
\end{equation}
We assume that the properties of each halo $i$ are defined solely by its mass $M_i$, such that $W_{{\rm u},i}= W_{\rm u}(M_i,r)$, and that the halo masses are distributed according to the halo-mass-distribution function $n(M)$, where $n(M)\,\diff M$ is the number density of haloes with masses between $M$ and $M+\diff M$.  With these assumptions we can find the mean value of $\theta_{\rm u}(\mathbf{x})$ by averaging over all haloes,
\begin{align}
\label{eq:meantheta}
\average{\hat{\theta}_{\rm u}(\mathbf{x})} = \left \langle \sum_i N_i\; W_{{\rm u}}(M_i,|\mathbf{x}-\mathbf{x}_i|) \right \rangle 
= \int_0^{\infty}  \diff M\; n(M) \int \diff^3 x'\; W_{\rm u}(M,|\mathbf{x}-\mathbf{x}'|)
=\int_0^{\infty}  \diff M\; W_{\rm u}(M) n(M)\;.
\end{align}
where we have translated the ensemble average, including $N_i$, into $\int  \diff M\; n(M)\; \Delta V_i$ with $\Delta V_i$ denoting the volume element.  The sum over $i$ can then be converted into an integral over the volume, $ \int \diff^3 x'$. To obtain the last equality in equation~(\ref{eq:meantheta}) we have separated the halo shape information through $W_{\rm u}(M,x)= W_{\rm u}(M) U_{\rm u}(M,x)$,  where $U_{\rm u}(M,x)$ is the normalised halo profile,
\begin{equation}
\int \diff^3 x\; U_{\rm u}(M,x) = 1\;.
\label{eq:normU}
\end{equation}
All the information about the amplitude of the halo profile is contained in $W_{\rm u}(M)$. Similarly we can define $\hat{W}_{\rm u}(M,k)= W_{\rm u}(M) \hat{U}_{\rm u}(M,k)$, where $\hat{U}_{\rm u}(M,k)$ is the Fourier transform of $ U_{\rm u}(M,x)$. From equation~(\ref{eq:normU}) we can conclude that $\hat{U}_{\rm u}(M,k\to0)=1$, which implies that $\hat{W}_{\rm u}(M,k\to0)=\hat{W}_{\rm u}(M)$. We can understand this result by considering that at large scales a halo acts as a point mass, which translates to a constant in Fourier space.  

Next we consider the correlation between two fields, $\theta_u$ and $\theta_v$, which could be identical, for example matter--matter, or different, for example matter--galaxies.  
Our fields are real and translationally invariant such that 
\begin{equation}
\average{\theta_{\rm u}(\bold{x})\;\theta_{\rm v}(\bold{x}')}=\xi_{\rm uv}(|\bold{x}-\bold{x}'|) \, ,
\label{eq:correlationuv}
\end{equation}
where $\xi_{\rm uv}$ is the two point correlation function between the two fields and it only depends on the separation $|\bold{x}-\bold{x}'|$. 
In Fourier space we have an equivalent relation for the power spectrum, $P_{\rm uv}(k)$,
\begin{equation}
\average{\hat{\theta}_{\rm u}(\bold{k})\hat{\theta}^*_{\rm v}(\bold{k}')}=(2\pi)^3 \delta_{\rm D}(\mathbf{k} -\mathbf{k}') P_{\rm uv}(k) \, ,
\label{eq:poweruv}
\end{equation}
where $\delta_{\rm D}$ is the Dirac delta function. The dimension of the power spectrum in equation~(\ref{eq:poweruv}) is volume times the dimensions of $\theta_{\rm u}$ and $\theta_{\rm v}$. Therefore we will sometimes use this definition of power spectrum instead
\begin{equation}
\Delta^2_{\rm uv}(k)= 4\pi\left(\frac{k}{2\pi}\right)^3 P_{\rm uv}(k)\ ,
\end{equation}
which removes the dependence on a volume dimension. 

We can find the power spectrum by inserting for the fields from equation~(\ref{eq:ffield}),
\begin{equation}
\average{\hat{\theta}_{\rm u}(\bold{k})\hat{\theta}^*_{\rm v}(\bold{k'})} = \left\langle\sum_{i,j} \mathrm{e}^{-{\rm i}\mathbf{k}\cdot\mathbf{x}_i}\; \mathrm{e}^{{\rm i}\mathbf{k}'\cdot\mathbf{x}_j}\; N_i\; N_j\; \hat{W}_{{\rm u},i}(M,k) \;\hat{W}_{{\rm v},j}(M,k') \right\rangle\ .
\label{eq:fullpuv}
\end{equation}
We can separate the sums above into two parts: when $i=j$, we measure field correlations within a single halo,  corresponding to the {\it one-halo} term,  $P^\mathrm{1h}_{\rm uv}(k)$;  when $i\neq j$, we measure field correlations between distinct haloes, called the {\it two-halo} term, $P^\mathrm{2h}_{\rm uv}(k)$.

\subsubsection{The one-halo term}
\label{sec:one_halo_term}
The one-halo term, where $i=j$ in equation~(\ref{eq:fullpuv}), is given by
\begin{equation}
\average{\hat{\theta}_{\rm u}(\bold{k})\hat{\theta}^*_{\rm v}(\bold{k'})} = \left\langle\sum_i \mathrm{e}^{-{\rm i}(\mathbf{k}-\mathbf{k}')\cdot\mathbf{x}_i} N_i\; \hat{W}_{{\rm u},i}(M,k)\; \hat{W}_{{\rm v},i}(M,k') \right\rangle\ ,
\label{eq:one_halo_term_sum}
\end{equation}
where we have used $N_i^2=N_i$.  We take similar steps to what was done in equation~(\ref{eq:meantheta}) to turn the ensemble average and the sum into continuous integrals,
\begin{align}
\average{\hat{\theta}_{\rm u}(\bold{k})\hat{\theta}^*_{\rm v}(\bold{k'})} = \int_0^{\infty}  \diff M\; n(M) \int \diff^3 x\; \mathrm{e}^{-{\rm i}(\mathbf{k}-\mathbf{k}')\cdot\mathbf{x}_i}\;  \hat{W}_{\rm u}(M,k)\;\hat{W}_{\rm v}(M,k')\;.
\label{eq:one_halo_term_theta}
\end{align}
The integral over the volume results in a $(2\pi)^3 \delta_{\rm D}(\mathbf{k}-\mathbf{k}')$.  Comparing equations~\eqref{eq:one_halo_term_theta} to \eqref{eq:poweruv} we arrive at the one-halo power spectrum,
\begin{equation}
P^\mathrm{1h}_{\rm uv}(k)=\int_0^\infty \hat{W}_{\rm u}(M,k)\;\hat{W}_{\rm v}(M,k) \; n(M)\,\diff M\ .
\label{eq:one_halo_term}
\end{equation}
The $k\to0$ limit of the one-halo term is independent of the shape of the halo profile as discussed after equation~(\ref{eq:normU}), $W_{\rm u}(M,k\to0)\to W_{\rm u}(M)$. For this reason at large scales the one-halo term only contributes as a constant $P(k)$, so-called shot noise. 

\subsubsection{The two-halo term}
\label{sec:two_halo_term}

The two-halo term is defined when $i\neq j$ in equation~(\ref{eq:fullpuv}).   This time to turn the sums and the ensemble average into integrals we need to also consider the correlation between the positions of haloes,  $\average{N_i\;N_j}=\xi_{\rm hh}^{ij}(|\bold{x}_i-\bold{x}_j|)$.  
We then follow the same steps as the one-halo term and obtain two integrals over halo masses and two over the volume elements where these haloes reside,
\begin{align}
\label{eq:two_halo_term_theta}
\average{\hat{\theta}_{\rm u}(\bold{k})\hat{\theta}^*_{\rm v}(\bold{k'})} &= \int_0^\infty  \diff M_1 \int_0^\infty\diff M_2\; n(M_1)\; n(M_2)\;\hat{W}_{\rm u}(M_1,k)\; \hat{W}_{\rm v}(M_2,k) \\ \nonumber
& \times \int \diff x^3 \int \diff x'^3 \;\mathrm{e}^{-{\rm i}\mathbf{k}\cdot\mathbf{x}}\; \mathrm{e}^{{\rm i}\mathbf{k}'\cdot\mathbf{x}'} \average{N(M_1,\mathbf{x})\;N(M_2,\mathbf{x}')}\,  .
\end{align}
We also know that
\begin{equation}
\int \diff x^3 \int \diff x'^3 \;\mathrm{e}^{-{\rm i}\mathbf{k}\cdot\mathbf{x}}\; \mathrm{e}^{{\rm i}\mathbf{k}'\cdot\mathbf{x}'} \average{N(M_1,\mathbf{x})\;N(M_2,\mathbf{x}')} = \average{\hat{N}(M_1,\mathbf{k})\;\hat{N}^*(M_2,\mathbf{k}')}= (2\pi)^3 \delta_{\rm D}(\mathbf{k}-\mathbf{k}') P_\mathrm{hh}(M_1,M_2,k)\, ,
\label{eq:hh_power}
\end{equation}
where $P_\mathrm{hh}(M_1,M_2,k)$ is the power spectrum of the halo centres with the shot-noise contribution subtracted.  Inserting for the two volume integrals in equation~(\ref{eq:two_halo_term_theta}) from equation~(\ref{eq:hh_power}) we find the two-halo power spectrum,
\begin{equation}
P^\mathrm{2h}_{\rm uv}(k)=\int_0^\infty\int_0^\infty\diff M_1\;\diff M_2\; P_\mathrm{hh}(M_1,M_2,k)\;
\hat{W}_{\rm u}(M_1,k)\;\hat{W}_{\rm v}(M_2,k)\;n(M_1)\;n(M_2)\ .
\label{eq:full_two_halo_term}
\end{equation}
As haloes are biased tracers of the underlying matter field, we can approximate the power spectrum of the halo centres as
\begin{equation}
P_\mathrm{hh}(M_1,M_2,k)=b(M_1)b(M_2)P^\mathrm{lin}_\mathrm{mm}(k)[1+\Bnl(M_1,M_2,k)]\ ,
\label{eq:Bnl_def}
\end{equation}
where $b(M)$ is the linear bias of haloes with mass $M$, and $P^\mathrm{lin}_\mathrm{mm}(k)$ is the linear-theory matter power spectrum.  The function $\Bnl$ then models all non-linear effects that are missing from the linear-bias--linear-field model,  vanishing on large-scales: $\Bnl(M_1,M_2,k\to0)=0$.  With this model we arrive at the two-halo power spectrum
\begin{equation}
P^\mathrm{2h}_{\rm uv}(k)=P^\mathrm{lin}_\mathrm{mm}(k)I^\mathrm{nl}_{\rm uv}(k)+
P^\mathrm{lin}_\mathrm{mm}(k)\prod_{n={\rm u,v}}\left[\int_0^\infty \hat{W}_n(M,k)b(M)n(M)\mathrm{d}M\right]\ ,
\label{eq:two_halo_term}
\end{equation}
where the non-linear halo bias modelling is captured in the term
\begin{equation}
I^{\mathrm{nl}}_{\rm uv}(k)=\int_0^\infty\int_0^\infty \Bnl(M_1,M_2,k)
\hat{W}_{\rm u}(M_1,k)\hat{W}_{\rm v}(M_2,k)b(M_1)b(M_2)n(M_1)n(M_2)\,\mathrm{d}M_1\mathrm{d}M_2\ .
\label{eq:Inl}
\end{equation}
It is common to set $I^\mathrm{nl}_{\rm uv}(k)=0$ for all $k$ and therefore assume that halo bias is linear,  as there is no analytical solution for this term.  We will discuss this term in more detail in section~\ref{sec:non_linear_bias}.

\subsection{Discrete tracers}
\label{sec:discrete_tracers}

With some small modifications, the theory described in the previous subsection can be applied to discrete tracers, such as galaxies. Modifications are necessary for two reasons: the first is that there can be a non-negligible scatter in the number of tracers that occupy haloes of the same mass; the second is that when computing the autocorrelation of a discrete tracer field, there is an automatic correlation of the field with itself at zero separation, the so-called shot noise. In configuration space this manifests at $r=0$ in the correlation function, but in Fourier space this is spread evenly over all wavenumbers, resulting in a constant shot noise power spectrum, $P^\mathrm{sn}=1/\bar{n}$, where $\bar{n}$ is the mean tracer number density.

In the following we consider a field of galaxies and in particular the galaxy number density contrast,  $\theta_{\rm u}= \delta_{\rm g} = (n_{\rm g}-\bar{n}_{\rm g})/\bar{n}_{\rm g}$,  although all calculations apply to any discrete tracer.  The mean number density of galaxies, $\bar{n}_{\rm g}$, is defined as
\begin{equation}
\bar{n}_{\rm g} = \average{n_{\rm g}} =  
\int_0^\infty \diff M \; n(M)\; N_{\rm g}(M) \int \diff^3 x \;U_{\rm g}(M,|\mathbf{x}-\mathbf{x}'|)= \int_0^\infty \diff M \; n(M)\; N_{\rm g}(M),
\label{eqn:ng}
\end{equation}
where we have followed the same logic as in equation~(\ref{eq:meantheta}). $N_{\rm g}(M)$ is the number of galaxies occupying a halo of mass $M$ and $U_\mathrm{g}(M,k)$ is the normalised distribution of the galaxies within the halo. 

Let us first assume that there is no scatter in the halo-occupation distribution (HOD), $N_{\rm g}(M)$.  The temptation when converting equations~(\ref{eq:one_halo_term}) and (\ref{eq:full_two_halo_term}) or (\ref{eq:two_halo_term}) to be appropriate for discrete tracers is to set
\begin{equation}
\hat{W}_\mathrm{g}(M,k) = \frac{N_\mathrm{g}(M)}{\bar{n}_\mathrm{g}}\hat{U}_\mathrm{g}(M,k)\ ,
\label{eq:discrete_substitution}
\end{equation}
%
This substitution works for the two-halo term, and for cross spectra of discrete tracer populations, but fails when computing the autospectrum of a discrete tracer field because it fails to take into account the fact that the field necessarily self correlates.  In this case $N_{\rm g}$ tracers create $N_{\rm g}$ contributions to the self correlation or the shot noise.  Importantly,  these shot noise terms are independent of the distribution of the tracers,  $U_{\rm g}$. To arrive at the correct one-halo expression,  we separate the cross galaxy term from the auto galaxy term by subtracting an $N_{\rm g}(M)\;U^2_{\rm g}(M)\;n(M)$ from the integrand and replacing it with the shot noise contribution, $N_{\rm g}(M)n(M)$\footnote{Another way to think of this is that if a halo contains only one galaxy, the one-halo term should be pure shot noise, with no dependence on the halo profile; the $N(N-1)$ term ensures the cancellation of this part of the one-halo term in that case.},
\begin{equation}
P^\mathrm{1h}_\mathrm{gg}(k)=\frac{1}{\bar{n}^2_\mathrm{g}}
\int_0^\infty \left[N_\mathrm{g}(M)(N_\mathrm{g}(M)-1)\hat{U}^2_\mathrm{g}(M,k)
+N_\mathrm{g}\right]n(M)\,\diff M\ ,
\label{eq:one_halo_term_discrete}
\end{equation}
this is often written as
\begin{equation}
P^\mathrm{1h}_\mathrm{gg}(k)=\frac{1}{\bar{n}^2_\mathrm{g}}
\int_0^\infty N_\mathrm{g}(M)(N_\mathrm{g}(M)-1)\hat{U}^2_\mathrm{g}(M,k) n(M)\,\diff M+P^\mathrm{sn}_\mathrm{gg}\ ,
\label{eq:one_halo_term_discrete_shotnoise}
\end{equation}
where
\begin{equation}
P^\mathrm{sn}_\mathrm{gg} = \frac{1}{\bar{n}^2_\mathrm{g}}
\int_0^\infty N_\mathrm{g}(M) n(M)\,\diff M\ = \frac{1}{\bar{n}_\mathrm{g}}\ .
\label{eq:shot_noise}
\end{equation}
This shot-noise term is often, but not always, subtracted from measured spectra, although it shows up again in the covariance matrix.  Note that in the $N\to\infty$ limit $N^2\gg N$ and equation~(\ref{eq:one_halo_term_discrete}) returns to the form obtainable from the substitution in equation~(\ref{eq:discrete_substitution}). This limit is appropriate for a general emissive profile, for example matter haloes taken to be composed from sub-atomic dark-matter particles, or even those from comparatively massive simulation particles. In this latter case, though the shot noise contribution is a real contribution, it is usually subtracted from simulation measurements because it is an artefact that arises due to the discretisation-techniques necessarily employed by \nbody simulations.

 In contrast, if we consider the cross spectrum between two different discrete populations ($\mathrm{g}$ and $\mathrm{g}'$; which may live in the same haloes), the one-halo term would be
\begin{equation}
 P^\mathrm{1h}_\mathrm{gg'}(k)=
 \frac{1}{\bar{n}_\mathrm{g}\bar{n}_\mathrm{g'}}
 \int_0^\infty N_\mathrm{g}(M)N_\mathrm{g'}(M)\hat{U}_\mathrm{g}(M,k)\hat{U}_\mathrm{g'}(M,k) n(M)\,\diff M\ ,
 \label{eq:one_halo_term_discrete_cross}
 \end{equation}
 which is exactly the generalization of the result from the continuous emissivity case. Note that evaluating equation~(\ref{eq:one_halo_term_discrete}) for haloes that contain $N_\mathrm{g}=1$ generates a pure shot-noise contribution. It is only when $N_\mathrm{g}>1$ that terms that involve the halo profile are generated, as one would expect.

When considering galaxy populations, it is usual that the galaxy population is broken down into separate contributions from central galaxies, with occupation number either $0$ or $1$, and satellite galaxies. It is also usual that some scatter is accounted for in the occupation numbers at fixed halo mass, which means we must keep the expectation value from equation~(\ref{eq:one_halo_term_sum}) to give:
\begin{equation}
P^\mathrm{1h}_\mathrm{cc}(k)=\frac{1}{\bar{n}^2_\mathrm{g}}
\int_0^\infty \average{N_\mathrm{c}(M)} n(M)\,\diff M\ ;
\label{eq:one_halo_central_central}
\end{equation}
\begin{equation}
P^\mathrm{1h}_\mathrm{ss}(k)=\frac{1}{\bar{n}^2_\mathrm{g}}
\int_0^\infty \left[\average{N_\mathrm{s}(M)(N_\mathrm{s}(M)-1)}\hat{U}^2_\mathrm{s}(M,k)+\average{N_\mathrm{s}(M)}\right] n(M)\,\diff M\ ;
\label{eq:one_halo_satellite_satellite}
\end{equation}
\begin{equation}
P^\mathrm{1h}_\mathrm{cs}(k)=
\frac{1}{\bar{n}^2_\mathrm{g}}
\int_0^\infty \average{N_\mathrm{c}(M)N_\mathrm{s}(M)}\hat{U}_\mathrm{c}(M,k)\hat{U}_\mathrm{s}(M,k) n(M)\,\diff M\ .
\label{eq:one_halo_central_satellite}
\end{equation}
We have used the fact that the occupation number of the central galaxies is $N_\mathrm{c}=0,1$ to eliminate the $\average{N_\mathrm{c}(N_\mathrm{c}-1)}$ term from equation~(\ref{eq:one_halo_central_central}), which leaves a pure shot-noise contribution. It is often taken that the central galaxies lie at the exact halo centre, in which case $\hat{U}_\mathrm{c}=1$, but we have left this term, which only enters the central--satellite cross spectrum, for completeness (some authors consider mis-centring, which can be included via this term). It is often taken that the central and satellite occupation numbers are independent: $\average{N_\mathrm{c}N_\mathrm{s}}=\average{N_\mathrm{c}}\average{N_\mathrm{s}}$ or otherwise the `central condition' is imposed such that satellites can only exist if there is a central galaxy: $\average{N_\mathrm{c}N_\mathrm{s}}=\average{N_\mathrm{s}}$. It is often assumed that the satellite occupation is determined by Poisson statistics: $\average{N_\mathrm{s}(N_\mathrm{s}-1)} = \average{N_\mathrm{s}}^2$. If shot-noise is subtracted this eliminates $P^\mathrm{1h}_\mathrm{cc}$ entirely, and also eliminates the term $\propto\average{N_\mathrm{s}}$ in the square bracket in equation~(\ref{eq:one_halo_satellite_satellite}); this is the most common form of these equations to be found in the literature. The full galaxy autospectrum can be constructed from these constituents via
\begin{equation}
P_\mathrm{gg}(k)=P_\mathrm{cc}(k)+2P_\mathrm{cs}(k)+P_\mathrm{ss}(k)\ .
\label{eq:galaxy_power_sum}
\end{equation}
Some expressions similar to equation~(\ref{eq:galaxy_power_sum}) contain pre-factors of the fraction of galaxies that are centrals or satellites. We avoid this here by defining the central and satellite fields as overdensity with respect to the total number of galaxies.
Note that it is \emph{not} possible to arrive at the same result for $P_\mathrm{gg}$ by replacing $N_\mathrm{g}\to N_\mathrm{c}+N_\mathrm{s}$ and $U_\mathrm{g}\to U_\mathrm{s}$ in equation~(\ref{eq:one_halo_term_discrete}) because this cannot account for the unique clustering and occupation properties of two distinct galaxy populations. A common \emph{approximate} \citep[\eg][]{Seljak2000} expression can be obtained by replacing the profile power of $2$ in equation~(\ref{eq:one_halo_term_discrete}) with $1$ if $\average{N_\mathrm{g}}\lesssim 2$ and retaining $2$ otherwise. This  roughly accounts for the fact that if more than one satellite is present the one-halo contribution to the power is dominated by the satellite auto correlation (equation~\ref{eq:one_halo_satellite_satellite}), whereas if a single satellite is present then it is dominated by the central--satellite cross correlation (equation~\ref{eq:one_halo_central_satellite}). Another reasonable approximation would be to use equation~(\ref{eq:one_halo_term_discrete}) with an occupation-number-weighted halo profile
\begin{equation}
\hat{U}_\mathrm{g}(M,k)\simeq
\frac{\average{N_\mathrm{c}}\hat{U}_\mathrm{c}(M,k)+\average{N_\mathrm{s}}\hat{U}_\mathrm{s}(M,k)}
{\average{N_\mathrm{c}}+\average{N_\mathrm{s}}}\ ,
\label{eq:approx_gal_window}
\end{equation}
together with
\begin{equation}
\average{N_\mathrm{g}(N_\mathrm{g}-1)}=\average{N_\mathrm{s}(N_\mathrm{s}-1)}+2\average{N_\mathrm{c}N_\mathrm{s}}\ .
\label{eq:Ng_exp}
\end{equation}
This last equation is always true as long as $\average{N_\mathrm{c}(N_\mathrm{c}-1)}=0$ (\ie $0$ or $1$ central galaxy only). The two terms on the right-hand side of equation~(\ref{eq:Ng_exp}) can be evaluated following the logic in the paragraph after equation~(\ref{eq:one_halo_central_satellite}). Using equation~(\ref{eq:approx_gal_window}) in equation~(\ref{eq:one_halo_term_discrete}) is not perfect, but the relative error can be small compared to the proper evaluation of equations~(\ref{eq:one_halo_central_central}--\ref{eq:galaxy_power_sum}) depending on the halo-occupation model.

\hrulefill
\vspace{0.5cm}
\twocolumngrid
\subsection{Matter}
\label{sec:matter}

Another special case is to evaluate the halo model solutions for the `matter' distribution in order to evaluate the matter power spectrum, but this involves some unique considerations. First, note that the adopted halo mass function and linear halo bias \emph{must} satisfy the following properties for any power spectrum involving the matter to have the correct large-scale limit
\begin{equation}
\int_0^\infty Mn(M)\,\mathrm{d}M=\bar\rho\ ,
\label{eq:mass_normalisation}
\end{equation}
\begin{equation}
\int_0^\infty Mb(M)n(M)\,\mathrm{d}M=\bar\rho\ .
\label{eq:bias_normalisation}
\end{equation}
where $\bar\rho$ is the mean comoving cosmological matter density. 
In other words, these equations enforce that all matter is contained in haloes and that, on average, matter is unbiased with respect to itself. Achieving these limits is difficult numerically because of the large amount of mass contained in low-mass haloes according to most popular mass functions.  Therefore, special care must be taken with the two-halo integral in the case of power spectra that involve the matter field to ensure that equation~(\ref{eq:bias_normalisation}) holds \citep[see appendix A of][]{Mead2020}.  Some popular mass function and bias relations enforce these consistency relations while others do not. If they do not, it might be possible to manually enforce these relations, but this normally involves manipulating the low-mass halo population.

\begin{figure}
\begin{center}
\includegraphics[width=\columnwidth]{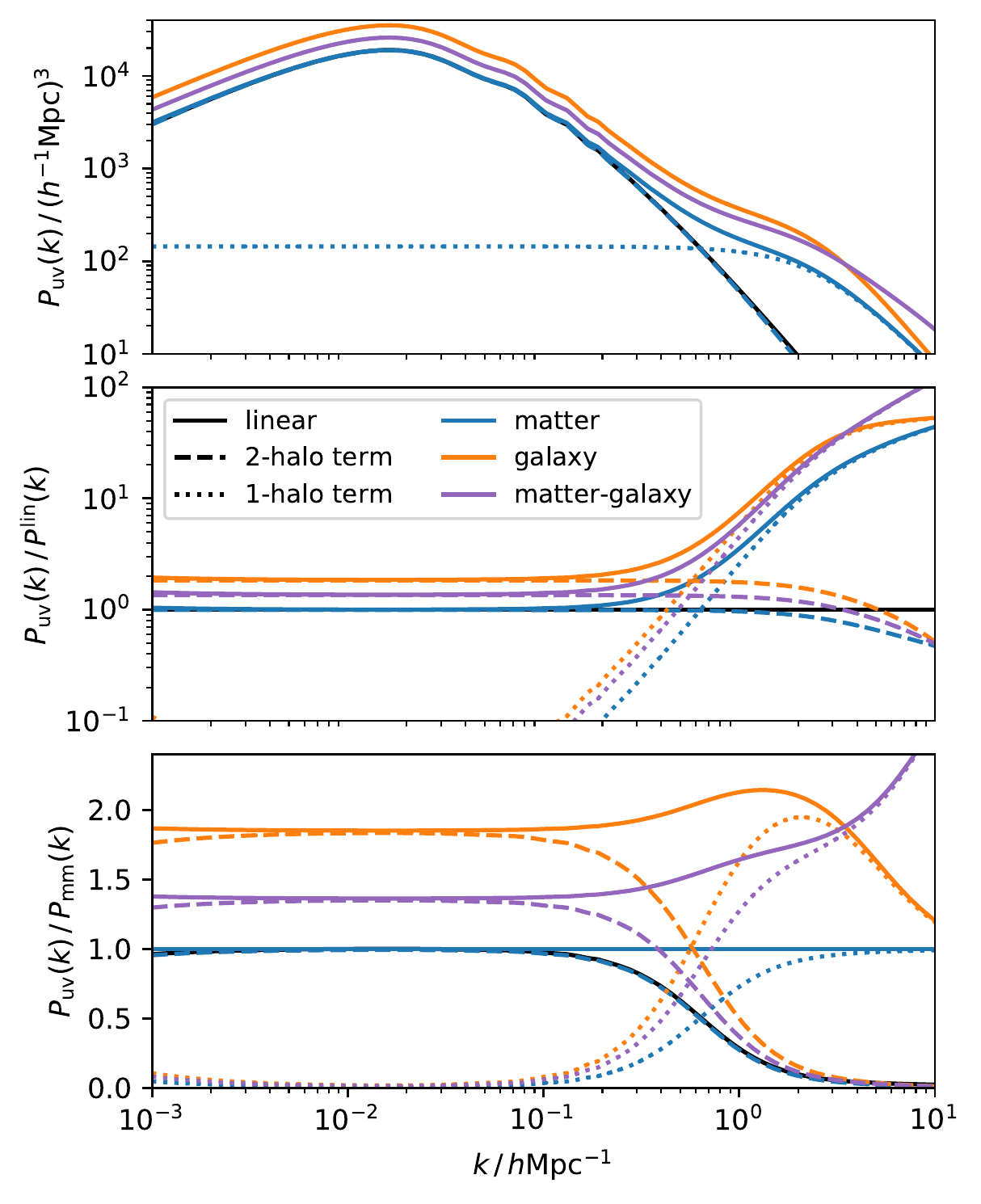}
\end{center}
\caption{The upper panel shows example power spectra computed using the halo model at $z=0$: matter, galaxy and matter--galaxy. The linear spectrum is shown for comparison as well as the breakdown into two- and one-halo terms where that does not clutter the plots. The middle panel shows the ratio of each spectrum to linear while the lower panel shows the ratio to the non-linear halo model matter spectrum. The galaxy sample can be seen to be positively biased relative to the matter. This bias is scale-independent at large scales, but becomes scale dependent at intermediate scales where the difference in the ways that galaxies and matter occupy haloes becomes important.}
\label{fig:power}
\end{figure}

In the special case of the power spectrum for matter density contrast, $\delta_{\rm m} = (\rho-\bar{\rho})/\bar{\rho}$,  we use $W_\mathrm{m}(M,k)=MU_\mathrm{m}(M,k)/\bar\rho$ and equation~(\ref{eq:one_halo_term}) becomes
\begin{equation}
P^\mathrm{1h}_\mathrm{mm}(k)=\frac{1}{\bar\rho^2}\int_0^\infty M^2\hat{U}^2_\mathrm{m}(M,k) n(M)\,\diff M\ .
\label{eq:one_halo_term_matter}
\end{equation}
while equation~(\ref{eq:two_halo_term}) becomes
\begin{align}
\begin{split}
P^\mathrm{2h}_\mathrm{mm}(k)&=P^\mathrm{lin}_\mathrm{mm}(k) \\
&\times\left[\frac{1}{\bar\rho}\int_0^\infty M \hat{U}_\mathrm{m}(M,k)b(M)n(M)\,\mathrm{d}M\right]^2\ ,
\end{split}
\end{align}
(ignoring non-linear halo biasing). We see that $P^\mathrm{2h}_\mathrm{uv}(k\to0)=P^\mathrm{lin}_\mathrm{mm}(k\to0)$ automatically as the term in the square brackets equals unity (equation~\ref{eq:bias_normalisation}) in this limit. For spectra other than matter this is no longer true, and in general the large-scale limit of the two halo term will be equal to the linear spectrum multiplied by amplitude factors (so-called bias) that account for the field content that arises from how the field populates haloes (e.g.  galaxy bias when the field is galaxy overdensity) and the halo bias.  Example power spectra at $z=0$ for a \LCDM model are shown in Fig.~\ref{fig:power} for matter, matter--galaxies and galaxies\footnote{Non-linear halo bias is ignored, the mass function is taken from \cite{Sheth1999}, halo concentration from \cite{Duffy2008} and the HOD from \cite{Zheng2005}; discussed in Sections~\ref{sec:halo_mass_function}, \ref{sec:dark_matter_haloes} and \ref{sec:HOD} respectively. The HOD parameters are $M_\mathrm{min}=M_0=M_1=10^{13}\Msun$, $\sigma_{\log_{10}M}=0.3$ and $\alpha=1$ (equations~\ref{eq:HOD_Nc} and \ref{eq:HOD_Ns}). Satellite galaxies are taken to trace matter. Shot noise is subtracted from the galaxy autospectra.}. The sample of galaxies chosen can be seen to be positively biased ($b\sim 1.3$) relative to the matter at large scales. At smaller scales the spectra have different shapes, a consequence of galaxy and matter occupying haloes in different ways.

\section{Selecting the ingredients}
\label{sec:ingredients}

When using the halo model it is necessary to make choices for the bias, halo mass function and halo profiles. Due to the lack of an analytical theory for non-linear gravitational clustering, it is common to calibrate these ingredients via \nbody simulations, or even via data. Within the halo model, haloes are treated as discrete entities, although real haloes never have clear boundaries. When defining haloes in simulations it is necessary to make a choice of boundary, and this choice must be consistent when using collections of simulation-calibrated ingredients within a halo model. The fundamental choice is how to identify a halo from the \nbody particle distribution. Two algorithms are in common usage: friends-of-friends (FoF; \citealt{Huchra1982}) and spherical-overdensity (SO; \citealt{Lacey1994}). 

The FoF scheme is simpler, with the only user-specified parameter being the `linking length', which defines the maximum distance between two particles that are considered to be part of the same halo.  All particles within the linking length of at least one other particle in the halo are joined to that halo. Typically the linking length is taken to be $b=0.2$ times the mean-inter-particle separation.  A FoF finder with this linking length applied to particles following an isothermal distribution ($\rho\propto r^{-2}$), will define a halo boundary such that the halo has a mean overdensity close to the analytical spherical-collapse result (see Section~\ref{sec:spherical_collapse}) in an Einstein-de Sitter model ($\Delta_\mathrm{v}\simeq178$ for $\Omega_{\rm m}=1$).

SO algorithms, on the other hand, first choose halo centres (usually from minima in the gravitational potential, but sometimes in the density) and then grow spheres out from these peaks until a fixed overdensity threshold has been reached. With SO there are several choices to be made: most obviously the value for the overdensity threshold ($200\bar\rho$ is common) but also exactly how to define the halo centres (how are continuous fields defined from the discrete particle distribution?) and how to count haloes as distinct entities (so-called percolation). While FoF is conceptually simpler, and is a mathematically unambiguous operation, SO is more common because it relates more closely to how halo formation is thought to occur and to how haloes are identified in data sets. SO haloes are, by definition, spherical (although the particle distribution that contributes to them may not be), whereas FoF haloes can be elongated structures, some of which may look to be distinct objects joined by a bridge. 

It should be noted that there is no single `correct' halo definition, and the best choice will depend on the observable that one is attempting to model. It is also important to be consistent, and to use sets of relations that have been calibrated on haloes identified using the same definition.  However, we note that it may be prudent to identify haloes using a cosmology-dependent definition, which accounts for the fact that halo formation happens at different rates, with different end results, in different cosmologies. Indeed, \cite{Courtin2011}, \cite{Despali2016} and \cite{Mead2017} have all noted that more `universal' (cosmology independent) behaviour is observed when haloes are identified with an overdensity threshold derived from the spherical-collapse model, with the general trend that haloes become denser the more dark energy takes hold of the expansion.  Useful fitting functions can be found in \cite{Bryan1998} and \cite{Mead2017}.

Aside from the virial radius, there are other physically motivated definitions of the halo boundary. 
For example,  the splashback radius which is defined as the largest distance in the orbit of particles accreted into the haloes and is usually measured by identifying the steepest gradient of the density profile \citep[for example][and references therein]{Fillmore1984,Bertschinger1985,Diemer2014, adhikari2014, More2015, Shi2016, Mansfield2017, Diemer2017,Oniel2021,Rana2023}. The value of the splashback radius is typical larger than or similar to the virial radius depending on the accretion rate of the halo.  
The turnaround radius is another typically larger physical radius which is motivated by the spherical collapse model and is defined as the distance at which particles reach zero velocity before falling into the halo \citep{Pavlidou2014,Tanoglidis2015,Korkidis2020}.  The turnaround radius is generally larger than the splashback radius and can be used as a test of gravity \citep[][]{Tanoglidis2015,Nojiri2018,Lopes2019,Capozziello2019,Pavlidou2020}.  In practice, the turnaround radius is defined as the radius at which the amplitude of the particle infall velocity is equal to the cosmic expansion.  
Two related and recent definitions for physical radii are given by \cite{Fong2021},  called the depletion radii.  The larger radius is defined as the radius at which the maximum depletion in the surrounding matter occurs and the smaller one as the radius with the maximum matter inflow into a halo.  In practice,  they are measured by finding the minimum of the ratio of halo-matter to matter-matter correlation functions and the minimum of the velocity profile around the halo  \citep[see also][]{Zhou2023, Gao2023}. 

\subsection{Preliminary definitions}
\label{sec:definitions}

Let us define a few useful quantities before we introduce the different ingredients.
The variance in the linear matter overdensity field when smoothed on comoving scale $R$ is
\begin{equation}
\sigma^2(R) =\int_0^\infty 4\pi \left(\frac{k}{2\pi}\right)^3 P^\mathrm{lin}(k)T^2(kR)\,\diff \ln k\ ,
\label{eq:sigmaR}
\end{equation}
where $T(kR)$ is the filter window function, which is almost exclusively taken to be a real-space top hat; the Fourier transform of which is
\begin{equation}
T(x)=\frac{3}{x^3}\left(\sin x - x\cos x\right)\ .
\end{equation}
The Lagrangian comoving scale,  $R$,  is the comoving radius of a sphere in a homogeneous Universe which contains a given mass of $M$,
\begin{equation}
M=\frac{4}{3}\pi R^3\bar\rho\ ,
\end{equation}
where $\bar\rho$ is the mean comoving matter density. This relation allows us to write $\sigma(R)$ in terms of mass. The `peak height', 
\begin{equation}
\nu(M)=\delta_\mathrm{c}/\sigma(M)\ ,
\label{eq:peak_height}
\end{equation}
is a useful quantity that increases monotonically with the halo mass.  Here $\delta_\mathrm{c}(z)\simeq 1.686/D(z)$ is the critical linear overdensity needed for haloes to collapse under the spherical-collapse model at redshift $z$ and $D(z)$ is the linear growth factor normalised to 1 at $z=0$.

\subsection{Halo mass function}
\label{sec:halo_mass_function}

\begin{table*}
\caption{A non-exhaustive list of popular halo mass functions, usually presented either as $f(\sigma)$ or $f(\nu)$. For each mass function we list the halo finder and definition, and we note that the mass function is a strong function of these choices: FoF haloes are uniquely defined by the linking length, but SO haloes need the halo-centre-finding and percolation scheme to be defined, as well as the overdensity threshold. SO haloes are then defined as spherical objects that are bounded such that they have a certain overdensity relative to either the background (\eg $200$) or critical (\eg $200$c) densities. Where a virial definition is used, this is most commonly evaluated using the fitting formula of \protect\cite{Bryan1998}. For every mass function, a linear halo bias can be derived using the peak-background split argument (equation~\ref{eq:peak_background_split}) but this may not be an accurate description of the large-scale bias \citep[\eg][]{Tinker2010, Manera2010}. The calibrated bias models of \protect\cite{Sheth2001} and \protect\cite{Tinker2010} are based on the mass functions of \protect\cite{Sheth1999} and \protect\cite{Tinker2008} respectively, but do not use the peak-background split. We also note whether each mass function is normalised such that all mass is in haloes (equation~\ref{eq:mass_normalisation}).}
\begin{center}
\begin{tabularx}{\textwidth}{lcccX}
\hline
Reference & Finder & Definition & Normalised & Notes \\
\hline
\cite{Press1974} & -- & -- & Yes & Purely analytical argument using the spherical collapse model, not connected to a specific mass definition, cosmology or redshift. \\
\cite{Sheth1999} & SO & virial & Yes & Original paper calculates the halo bias via the peak-background split; \cite*{Sheth2001} use an ellipsoidal-collapse argument for a more accurate bias. Cosmology dependence of $\delta_\mathrm{c}$ accounted for via spherical collapse. \\
\cite{Jenkins2001} & FoF & $0.2$ & No & First accurate parameterisation for FoF haloes.\\
\cite{Warren2006} & FoF & $0.2$ & No & Argument presented for resolution-dependent conversion between FoF and SO masses. Correction to FoF masses for low-particle haloes. \\
\cite{Reed2007} & FoF & $0.2$ & No & Depends on effective power spectrum index at the collapse scale, $n_\mathrm{eff}$, as well as $\nu$.\\
\cite{Peacock2007} & FoF & $0.2$ & Yes & Based on fit to model of \cite{Warren2006}. \\
\cite{Tinker2008} & SO & $200$--$3200$ & Both & Parametrised in terms of $\sigma$. Principle result is unnormalised and has redshift-dependent (non-universal) parameters for $\Delta_\mathrm{h}=200$. However, redshift-independent results are presented for a variety of other SO halo definitions, and these can be interpolated between (appendix B) for a virial halo definition. A normalised mass function is also presented (appendix C). \\
\cite{Tinker2010} & SO & $200$--$3200$ & Yes & Mass function is the same as the normalised (appendix C) version from \cite{Tinker2008} but recast in terms of $\nu$. Once again, redshift-dependent parameters are presented only for $\Delta_\mathrm{h}=200$. A calibrated halo bias is presented that fulfils equation~(\ref{eq:bias_normalisation}) without using the peak-background split argument. \\
\cite{Crocce2010} & FoF & $0.2$ & No & Uses functional form of \cite{Warren2006}. \\
\cite{Bhattacharya2011} & FoF & $0.2$ & No & Consider $w$CDM dark energy. \\
\cite{Courtin2011} & FoF & virial & No & Results demonstrate that virial definitions are `more universal'; semi-analytical relation to convert $\Delta_\mathrm{h}$ to FoF linking length. \\
\cite{Watson2013} & FoF/SO & various & No & Consider wide redshift range, from $z=30$ to $0$.\\
\cite{Despali2016} & SO & virial & No & Argues that the mass function is `more universal' when a cosmology-dependent virial criterion is used to identify haloes, rather than a fixed overdensity threshold. \\
\cite{DelPopolo2017} & -- & -- & Yes & Uses the excursion set approach and provides a semi-analytic solution with a mass dependent collapse threshold.   \\
\cite{McClintock2019a} & SO & $200$ & Yes & Parameters of a universal, $\sigma$-dependent, fitting function are emulated to encompass cosmology and redshift dependence. \\
\cite{Bocquet2020} & SO & $200$c & No & Mass function principle components are directly emulated. 
\\
\hline
\end{tabularx}
\end{center}
\label{tab:massfunctions}
\end{table*}

\begin{figure}
\begin{center}
\includegraphics[width=\columnwidth]{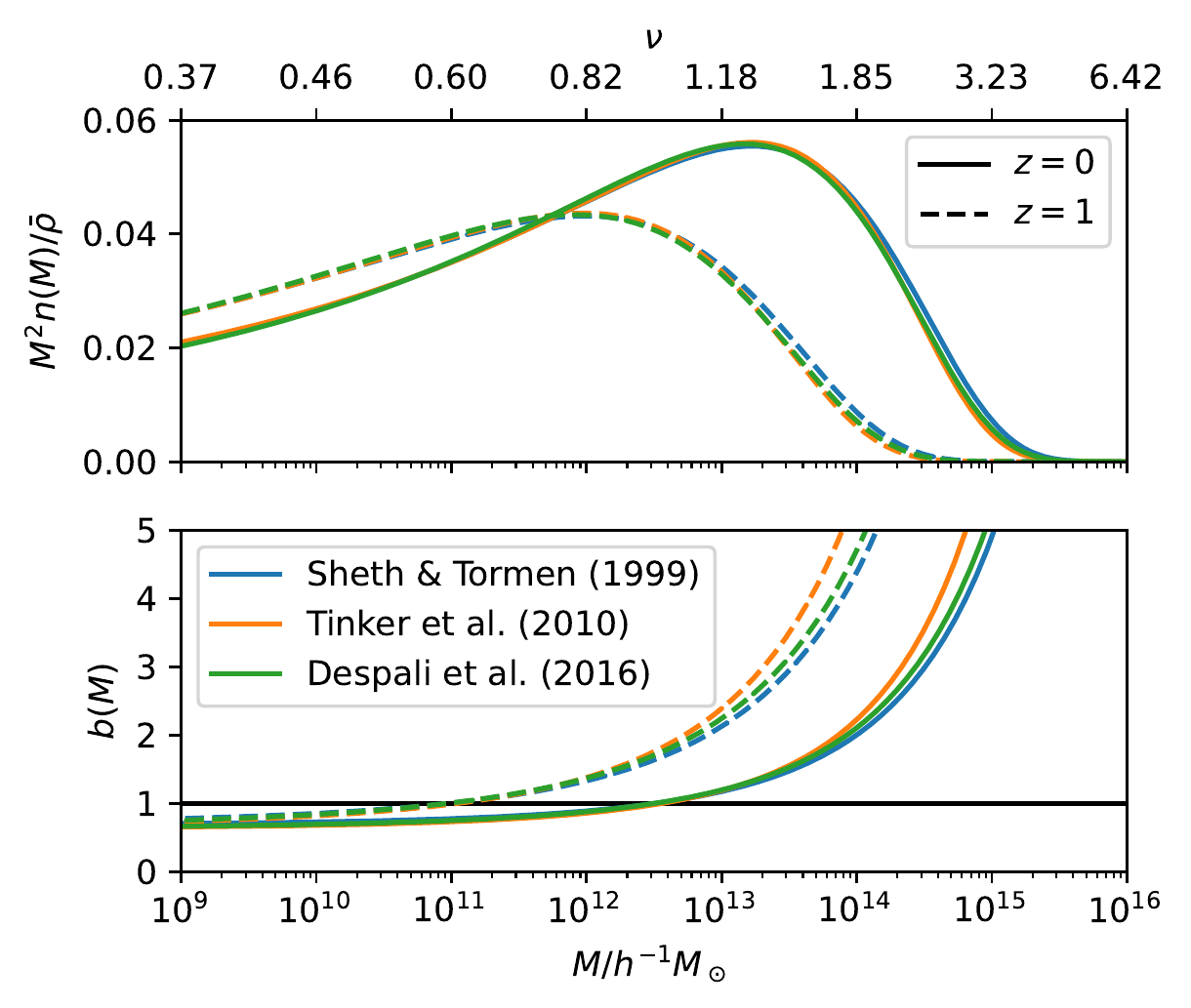}
\end{center}
\caption{Upper panel: Dimensionless multiplicity function, $M^2 n(M)/\bar\rho$, as a function of halo mass for three popular halo-mass functions \protect\citep{Sheth1999, Tinker2010, Despali2016} at $z=0$ (solid) and $z=1$ (dashed) for the virial halo definition. The top axis shows the $\nu$ values corresponding to halo mass at $z=0$ \emph{only} (the mapping is $z$-dependent). For normalised mass functions, integrating the multiplicity function over $\ln M$ will equal unity, and therefore the shape of the function determines the contribution of haloes in a logarithmic mass range to the total mass in the cosmos, with the peak determining those most important. As the universe evolves, more haloes of higher mass are created, but this is done at the expense of those of lower mass via mergers. Lower panel: the associated linear halo bias, low mass haloes are anti-biased ($0<b<1$) with a constant asymptotic value at low mass. The transition to biased, $b>1$, objects occurs around the non-linear mass ($\nu=1$). At fixed halo mass, haloes are comparatively rarer, and more highly biased, at higher $z$. The bias shown for \citecaption{Tinker2010} is from their calibrated fitting function, whereas in the other two cases it is from the peak-background split. \citecaption{Tinker2010} predicts fewer, but more highly biased, high-mass haloes compared to the other models.}
\label{fig:multiplicity}
\end{figure}

The halo mass function is usually parametrised in terms of either $\sigma$ or $\nu$, rather than $M$ directly, because it has been shown that the halo mass function (and also bias) exhibit close-to-universal behaviour as a function of cosmology and redshift in terms of these variables \citep[\eg][]{Press1974, Bond1991, Sheth1999, Tinker2008}. Analytical approaches for calculating (approximately) the halo mass function rely on either peaks theory or excursion sets. These methods start from the initial matter density field and relate some of its properties to the haloes that form later. For peaks theory, the focus is on peaks in the primordial matter density field \citep{Bardeen1986}, while excursion sets look at overdense regions \citep[\eg][]{Bond1991, Bond1996, Stein2019}.  The peak height, $\nu$, has been shown to be the relevant quantity to consider when calculating halo formation via peaks theory and excursion sets. For reference, for a vanilla \LCDM cosmology at $z=0$, $\nu=0.5$, $1$, $2$, and $3$ correspond to $\simeq 10^{10.4}$, $10^{12.5}$, $10^{14.2}$, and $10^{14.9}\Msun$. Some common mass functions are shown in Fig.~\ref{fig:multiplicity}, and the (generally non-linear) mapping between $M$ and $\nu$ can be read off the axes.

Now we can write the halo mass function in terms of the peak height,  by defining 
\begin{equation}
f(\nu)\,\diff\nu=\frac{M}{\bar\rho}n(M)\,\diff M\ .
\label{eq:nu_M_conversion}
\end{equation}
If all mass is to be contained in haloes, then $f(\nu)$ integrated over all $\nu\in[0,\infty]$ should equal unity, which derives from mass conservation (equation~\ref{eq:mass_normalisation}). Note that this condition is only imposed on some fitting functions. Other than the constraint imposed by mass conservation, the shape of the low-mass end of the halo-mass function is difficult to access through \nbody simulations due to finite particle resolution. Commonly-used fitting functions should be interpreted with caution in this regime. A common form of $f(\nu)$ to be found in the literature is that of \cite{Sheth1999}:
\begin{equation}
f(\nu)=A\left[1+(q\nu^2)^{-p}\right]\mathrm{e}^{-q\nu^2/2}\ ,
\label{eq:ST_massfunction}
\end{equation}
where $p$, $q$ and (sometimes) $A$ are fitted to simulated data. If $A$ is fitted independently of $p$ and $q$ then the mass function will not be normalised. If $A$ is not fitted then it depends on $p$ and $q$ via the normalisation condition. In Table~\ref{tab:massfunctions} we list some popular mass functions together with the halo finder and definition on which they were calibrated.

Finally we note that $\delta_\mathrm{c}\simeq 1.686$ is usually assumed; a value that corresponds to a universe with an Einstein-de Sitter background\footnote{A spatially flat cosmology with the matter density parameter $\Omega_{\rm m}=1$.}.  
Although,  $\delta_\mathrm{c}$ has a weak cosmology dependence \citep[\eg][]{Lacey1993} that can be calculated using the spherical-collapse model (fitting formulae: \citealt{Nakamura1997, Mead2017}). This cosmology dependence is often ignored in the conversion between $\nu$ and $M$. However, the general exponential form of the halo mass functions (\eg equation~\ref{eq:ST_massfunction}) can make this weak dependence have a larger impact than one might first assume.  For example,  spherical collapse predicts that $\delta_\mathrm{c}\simeq 1.676$ for $\Om=0.3$ \LCDM, a small decrease from the canonical $1.686$. However, this small difference results in a $\sim4$ per cent \emph{increase} in the abundance of rare ($\nu=4$; $M\simeq\sform{2}{15}\Msun$; $z=0$) haloes if the mass function of \cite{Sheth1999} applies. \cite{Courtin2011} and \cite{Mead2017} have suggested that retaining this cosmology dependence of $\delta_\mathrm{c}$ may improve the cosmological universality of halo mass functions and halo-model calculations.

\subsection{Linear halo bias}
\label{sec:lin_bias}

On scales large enough to comfortably encompass the largest haloes, the overdensity of haloes of any mass can be approximated by the (unconditional) halo mass function via the peak-background split argument \citep{Cole1989, Mo1996, Sheth2001}. The density field is thought of as a sum of large- and small-scale waves with haloes forming at global peaks; more peaks are pushed over the formation threshold when large and small-scale waves constructively interfere, which will be in regions of large-scale overdensity, leading to biased clustering. The peak-background split argument can be used to calculate an approximate linear halo bias from any mass function:
\begin{equation}
b(\nu) = 1-\frac{1}{\delta_\mathrm{c}}\left[1+\frac{\diff\ln f(\nu)}{\diff\ln\nu} \right]\ .
\label{eq:peak_background_split}
\end{equation}
If a particular $f(\nu)$ satisfies the mass-normalisation condition in equation~(\ref{eq:mass_normalisation}), then the combination of $f(\nu)$ with $b(\nu)$ calculated this way automatically satisfies the bias-normalisation condition in equation~(\ref{eq:bias_normalisation}). However, if these normalisation conditions are not important, then equation~(\ref{eq:peak_background_split}) can be still applied to any mass function to get an expression for the bias (although it may not be accurate). In practice, not satisfying the bias-normalisation condition is only a fundamental problem when calculating the matter spectrum, and only then if one explicitly integrates over all halo masses, which is not normally done in halo-model codes\footnote{See the discussion in Appendix A of \cite{Mead2020}}.

Note that the peak-background split is \emph{not} the only way to satisfy the bias normalisation condition, and popular bias relations (\citealt*{Sheth2001}; \citealt{Tinker2010}) satisfy the normalisation condition using other schemes. The accuracy of the peak-background split has been disputed \citep[\eg][]{Manera2010} and calibrated bias relations may therefore be preferred.

\subsection{Non-linear halo bias}
\label{sec:non_linear_bias}

As discussed in Subsection~\ref{sec:two_halo_term}, the beyond-linear portion of the halo bias is not often considered in halo-model calculations: It is common to set $\beta^\mathrm{nl}=0$ in equation~(\ref{eq:Bnl_def}), and therefore implicitly $I^\mathrm{nl}=0$ in equation~(\ref{eq:Inl}). This means that the `standard' two-halo term at large scales for any tracer is always the linear power multiplied by some scaling (bias) factors, which arise jointly through the linear halo bias and halo occupation. As shown by \cite{Mead2021b}, the lack of beyond-linear bias is mainly responsible for the poor performance of the standard halo model in the transition region between the two- and one-halo terms. At smaller scales the standard two-halo term is suppressed by the halo window functions, but this effect is often not visible in the total halo-model spectrum since it occurs on scales where the one-halo term tends to dominate the power. Note well that the presence of the window functions in the standard two-halo term is \emph{not} accounting for halo exclusion (the fact that spatially-exclusive haloes should not overlap), but instead is a blurring of correlation between points in different haloes caused by the fact that these points are not at the exact halo centre. In reality, the two-halo term should be further suppressed by halo exclusion (Section~\ref{sec:exclusion}), which is also absent in simple linear bias models.

\begin{figure}
\begin{center}
\includegraphics[width=\columnwidth]{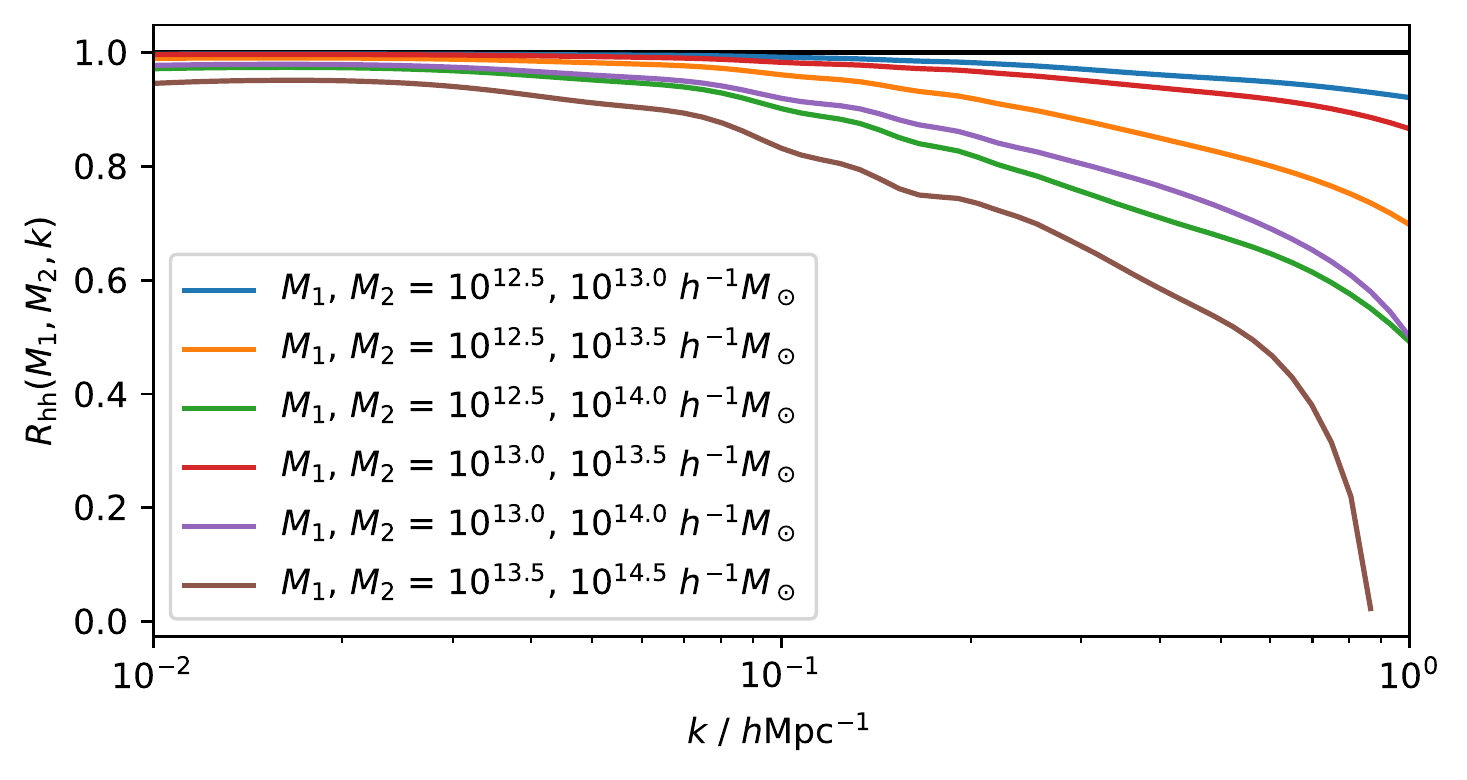}
\end{center}
\caption{Halo--halo correlation coefficient as a function of scale taken from the \DQ emulator. At large scales this is consistent with unity, but it drops below unity at smaller scales. This implies that there is a non-zero covariance in the clustering between haloes in different mass bins, which cannot be captured by a simple linear bias model.}
\label{fig:Rhh}
\end{figure}

The two-halo term accounts for inter-halo clustering and therefore the fundamentally-correct spectrum to include within the two-halo term is the halo power spectrum, for which no fitting function exists in the literature. In our notation (equation~\ref{eq:two_halo_term}) the beyond-linear portion of this is factored out into $\beta^\mathrm{nl}$. To illustrate some properties of this, we define the halo--halo cross correlation in Fourier space as 
\begin{align}
\begin{split}
R_\mathrm{hh}(M_1, &M_2, k)= \\
&\frac{P_\mathrm{hh}(M_1, M_2, k)}{\sqrt{P_\mathrm{hh}(M_1, M_1, k) P_\mathrm{hh}(M_2, M_2, k)}}\ ,
\label{eq:cross_correlation_coefficient}
\end{split}
\end{align}
and show this for various halo masses in Fig.~\ref{fig:Rhh}, where the correlation is calculated using the \DQ emulator of \cite{Nishimichi2019, Miyatake2022a}. The fact that this departs from unity at small scales indicates a non-zero covariance between the clustering of haloes in different mass bins \citep[][]{Hamaus2010, Baldauf2013, Schmidt2016}. This indicates that any model for the non-linear halo bias where the bias is separable,  
\begin{equation}
P_\mathrm{hh}(M_1, M_2, k)\simeq b(M_1, k)b(M_2, k) P_\mathrm{mm}(k)
\end{equation}
fails to describe the covariant structure, even if a scale dependent $b(M, k)$ \citep[\eg][]{Fedeli2014b} or if a non-linear $P_\mathrm{mm}(k)$ is used: the non-linear halo-bias is fundamentally a non-separable function! The structure of the halo power spectrum ensures that the clustering of haloes in one mass bin respond to the clustering of haloes in other mass bins. \cite{Tinker2005} use a fitting function in real space for the radial dependence of the non-linear halo bias and compute that as a function of the non-linear matter correlation function. This fitting function has no dependence on halo masses other than through the linear halo bias, and is thus difficult to interpret for different galaxy populations that may exist in very different haloes. 

Some authors \citep[\eg][]{Cacciato2012, vandenBosch2013}, particularly those interested in using the halo model to compute galaxy spectra, replace the linear spectrum that appears in equation~(\ref{eq:two_halo_term}) with the full non-linear matter spectrum\footnote{Usually from a fitting function, for example \halofit \citep{Smith2003, Takahashi2012} or \hmcode \citep{Mead2015b, Mead2021a}, although this could also come from an emulator \citep[\eg][]{Lawrence2017, Knabenhans2019, Angulo2020}.}. We note that this is inconsistent with the halo-model ethos, since in principle the non-linear matter power should be computable via the halo model. However, for galaxies, it has been demonstrated that using the non-linear power provides a better approximation at quasi-linear ($k\simeq0.1\iMpc$) scales compared to using the linear power. Despite this, we advise an abundance of caution: the non-linear matter power contains its own one-halo term, which arises due to the auto-convolution of the matter profiles at small scales. This feature has no analogue in the halo spectrum (which is dominated by exclusion at such scales), even though the shape of the halo spectrum may be super-linear at quasi-linear scales. This means that using the non-linear matter spectrum can be extremely wrong at small scales, and one virtue of the linear approximation is that it will be significantly less wrong. Incorporating a small-scale halo-exclusion model may ameliorate the problem induced when using the non-linear matter spectrum, with the `exclusion' term performing the joint job of dampening the excess one-halo power and accounting for the genuine spatially exclusivity of haloes, but does not escape the physical incorrectness of employing the non-linear matter spectrum in this role. Finally, the replacement $P^\mathrm{lin}(k)\to P^\mathrm{nl}(k)$, with $I^\mathrm{nl}(k)=0$, in equation~(\ref{eq:two_halo_term}) suffers from the same problem demonstrated in Fig.~\ref{fig:Rhh}; the value of $R_\mathrm{hh}=1$ always, contrary to measurements and theoretical expectations. 

Few authors have tackled the issue of non-linear halo bias in detail. \cite{Smith2007, Ginzburg2017} used the combination of perturbation theory for both the matter field and halo bias to demonstrate that improved predictions could be made for quasi-linear scales. Unfortunately, because perturbation theory fails at smaller scales, so does this method. \cite{Nishimichi2019} emulated the halo power spectrum directly, so that it could be incorporated within halo-model calculations of the galaxy power spectrum. This emulated power contains both classical non-linearity in the bias together with halo exclusion, since both effects are present in simulations and in the measured halo overdensity fields. Finally, \cite{Mead2021b} showed that incorporating the non-linear halo bias (measured from \nbody simulations) within halo-model calculations dramatically improves accuracy in the transition region. The $\beta^\mathrm{nl}$ defined in that work (equation~\ref{eq:Bnl_def}) is related to the halo stochasticity matrix defined by \cite{Hamaus2010} and the halo stochasticity covariance defined by \cite{Schmidt2016}.


\subsection{Dark matter halo profiles}
\label{sec:dark_matter_haloes}

By far the most common form taken for the density profile of collisionless matter is that of \citeauthor*{Navarro1997} (NFW; \citeyear{Navarro1997})
\begin{equation}
\rho(r)=\frac{\rho_\mathrm{s}}{r/r_\mathrm{s}(1+r/r_\mathrm{s})^2}\ ,
\label{eq:NFW}
\end{equation}
where $\rho_\mathrm{s}$ and $r_\mathrm{s}$ are the scale radius and density, both of which depend on the halo mass. The profile is usually truncated\footnote{The profile truncation can be sharp or smooth.  A smooth truncation for galaxy clusters has been shown to provide a better match to N-body simulations \citep[see for example][]{Oguri2011,Diemer2014}. } at the halo radius $r_\mathrm{h}$ and if this truncation is not imposed then it should be noted that the total mass of the profile is formally infinite. The halo radius (which need not necessarily be the `virial' radius\footnote{In the context of halo definitions, the `virial' radius is often used interchangeably with `halo' radius. It need not have anything to do with virialised haloes or the virial theorem.}) is calculated via
\begin{equation}
M=\frac{4}{3}\pi r_\mathrm{h}^3\Delta_\mathrm{h}\bar\rho\ ,
\label{eq:virial_radius}
\end{equation}
where $\Delta_\mathrm{h}$ is the halo overdensity with respect to the background matter density (usually either $200$, or $200$ times the critical density, or else the virial definition). 
A less common, but possibly more accurate, choice for the halo profile is that of \cite{Einasto1984}, 
\begin{equation}
\rho(r)=\rho_{\rm s}{\rm exp}\left(- \frac{2}{\alpha} \left[\frac{r}{r_{\rm s}} -1 \right]^\alpha\right) \ ,
\end{equation}
which has an extra `shape' parameter, $\alpha$, as well as an $r_\mathrm{s}$ similar to the NFW profile, \citep[also see][for more details]{Navarro2004,Gao2008}.
A recent update to the \cite{Einasto1984} halo profiles has been given by \cite{Diemer2022}, which includes two characteristic scales and fits to numerical simulations more accurately.  

\begin{table*}
\caption{A non-exhaustive list of popular NFW profile concentration--mass relations that have been fitted to data from \nbody simulations. Note that different samples of haloes may have been used in fitting each relation (\eg relaxed vs. all haloes) and different criteria may be employed to isolate unique halo centres, and different cosmologies may have been considered. We encourage the reader to carefully read the papers below to ensure that they understand the details of the relation they are using and to ensure that it is appropriate for their use case.}
\begin{center}
\begin{tabularx}{\textwidth}{lcX}
\hline
Reference & Definition & Notes \\
\hline
\cite{Navarro1997} & $200$c & Depends on a cosmology-dependent halo-collapse redshift that is calculated semi-analytically. \\
\cite{Bullock2001} & virial & Two relations presented in paper: a simple model where $c$ is a power-law in $M$ (although scaled by a cosmology-dependent non-linear mass) and a more complicated model where $c$ is related to a cosmology-dependent halo formation redshift, which is calculated semi-analytically. \\
\cite{Eke2001} & virial & Depends on a cosmology-dependent halo-collapse redshift that is calculated semi-analytically. \\
\cite{Neto2007} & $200$c & Only considered the Millennium \citet{Springel2005a} cosmology at $z=0$. \\
\cite{Maccio2008} & virial & Modified version of the \cite{Bullock2001} algorithm. \\
\cite{Duffy2008} & $200$, $200$c, virial & Simple $c(M)$ power-law relations are presented that are fitted to simulations of \wmap5 cosmology.  Explicit $z$ dependence. Separate relations for `relaxed' and `full' samples of haloes. \\
\cite{Prada2012} & $200$c & `Cosmology dependent' relation presented as a function of $\sigma(M, a)$. Upturn in halo concentration for high-mass haloes. \\
\cite{Kwan2013} & $200$c & Emulated relation for a variety of $w$CDM cosmologies. \\ 
\cite{Ludlow2014} & $200$c & Relates halo concentration to mass-accretion history. \\
\cite{Klypin2014} & $200$c & Parametrised in terms of $\nu$. \\
\cite{Diemer2015} & $200$c & Present a semi-analytical, cosmology-dependent model parametrised in terms of $\nu$ and $n_\mathrm{eff}$ -- the effective slope of the power spectrum on collapse scales. Demonstrates that concentration--mass relation is `most universal' when masses are defined via $200$c. \\
\cite{Correa2015} & $200$c & Relates halo concentration to mass-accretion history. Only applies to relaxed haloes. \\
\cite{Okoli2016} & $200$c & Focusses on relaxed low-mass haloes using analytical arguments. Cosmology dependence incorporated via $\nu$ dependence. \\
\cite{Ludlow2016} & $200$c & Applies for WDM as well as for CDM cosmologies. Depends on a collapse redshift that is calculated semi-analytically. \\
\cite{Child2018} & $200$c & Power-law relation but scaled via the cosmology-dependent non-linear mass. Also consider Einasto profiles. Individual and stacked halo profiles considered separately. \\
\cite{Diemer2019} & $200$c & Improved version of \cite{Diemer2015} with additional dependence on the logarithmic linear growth rate to capture non-standard expansion histories. \\
 \cite{Ishiyama2021} & $200$c, virial &  Uses the same functional form as \cite{Diemer2019}  but fitted to a larger simulation resulting in up to $5\%$ errors for a wide range of masses and redshifts.  \\
\hline
\end{tabularx}
\end{center}
\label{tab:concentration}
\end{table*}

To fully specify the halo profile in equation~(\ref{eq:NFW}) we need to know the scale radius, $r_\mathrm{s}$, which is usually related to $r_\mathrm{h}$ via a concentration--mass relation: $c=r_\mathrm{h}/r_\mathrm{s}$. These are always calibrated to haloes measured in \nbody simulations, and once again we stress that the relations will depend on the halo definition, as well as the details of precisely how the concentration was inferred from the measured halo sample. For example: Is the relation fitted to the mean or median halo profile in a mass bin, or to individual haloes? Is the cumulative profile fitted or the raw density? Are certain haloes discarded from the sample? Is the concentration inferred from the circular velocity profile? In Table~\ref{tab:concentration} we list some concentration--mass relations that are in common usage. We also note that a scatter in the concentration parameter at fixed halo mass is seen in haloes identified in \nbody simulations \citep[\eg][]{Jing2000, Bullock2001} with an approximate log-normal distribution with $\sigma_{\ln c}\simeq 0.3$. This scatter can be included in halo-model calculations (see Section~\ref{sec:scatter}).


The constant of proportionality from equation~(\ref{eq:NFW}) can be found by ensuring that integrating the density profile over the halo volume gives the correct enclosed mass:
\begin{equation}
\rho_\mathrm{s} = \frac{M}{4\pi r_\mathrm{h}^3}\left[\frac{c^3}{\ln(1+c)-c/(1+c)}\right]\ .
\end{equation}
In passing, we note that the mass enclosed at a given radius by an NFW profile is
\begin{equation}
M(r) = M\left[\frac{\ln(1+r/r_\mathrm{s})-(r/r_\mathrm{s})/(1+r/r_\mathrm{s})}{\ln(1+c)-c/(1+c)}\right]\ .
\end{equation}
%

\begin{figure}
\begin{center}
\includegraphics[width=\columnwidth]{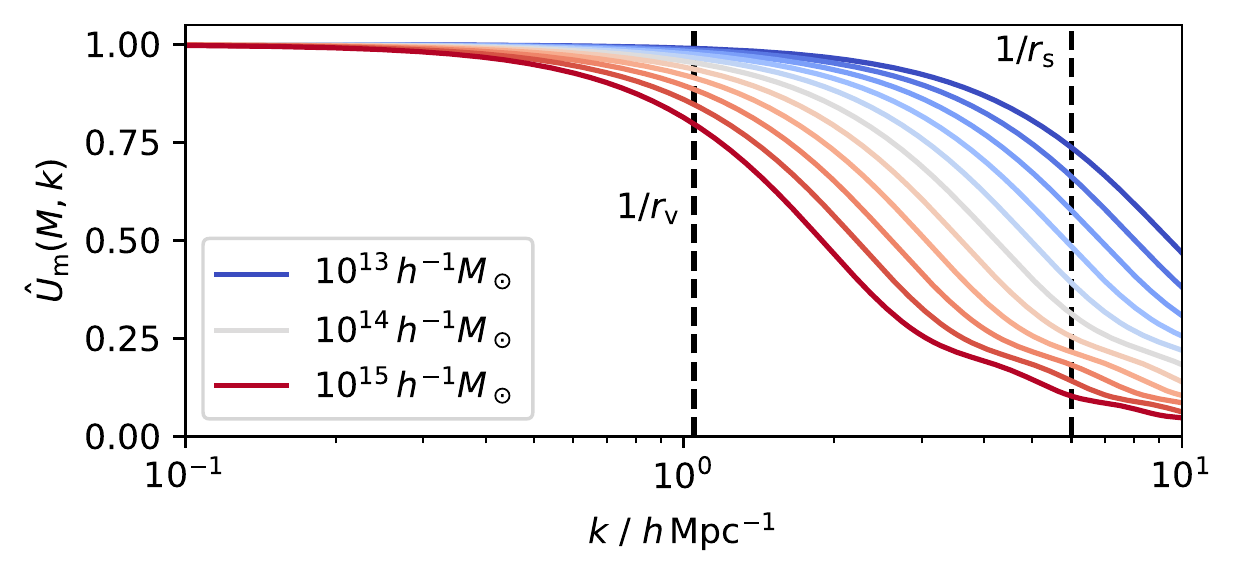}
\end{center}
\caption{NFW profile normalised Fourier transforms. We show a range of halo masses log-spanning $10^{13}$ to $10^{15}\Msun$ with parameters corresponding to a \LCDM cosmology at $z=0$. This takes in a range of virial radii from $0.44$ to $2.1\Mpc$ and concentrations from $6.9$ to $4.7$, with central $M=10^{14}\Msun$, $r_\mathrm{h}=0.95\Mpc$ and $c=5.7$. The vertical lines show approximate wavenumbers corresponding to the central halo virial and scale radius.}
\label{fig:window}
\end{figure}

Once the real-space halo profile has been specified it must be Fourier transformed (equation~\ref{eq:window_function}) for use in the power spectrum calculation (equations~\ref{eq:one_halo_term} and \ref{eq:two_halo_term}). Example normalised profile Fourier transforms are shown in Fig.~\ref{fig:window}. Note that the shape dependence of smaller haloes only affects the power at higher $k$. In the power spectrum calculation the normalised windows are multiplied by halo mass, which boosts the contribution from higher-mass haloes, but are also multiplied by the mass function, which reduces the contribution.

The choice of halo overdensity to use in the halo model, $\Delta_\mathrm{h}$, is fixed if using ingredients that have been calibrated on haloes identified via a SO finder, but it is less obvious what to choose for $\Delta_\mathrm{h}$ when dealing with FoF-identified haloes. Many schemes have been proposed to relate a linking length, $b$, to $\Delta_\mathrm{h}$, the simplest involve imagining halo profiles sampled by discrete particles, and then calculating the corresponding linking length at the halo boundary. This has the unattractive property that linking length depends on the halo profile \citep[\eg][]{Lukic2009}, so often a simple isothermal halo is assumed when performing this conversion, which leads to the approximate relation \citep{Lacey1994}:
\begin{equation}
\left(\frac{b}{0.2}\right)^{-3} \simeq \frac{\Delta_\mathrm{h}}{180}\ .
\end{equation}
The correspondence with the matter dominated spherical-collapse result is precisely why the linking length $b=0.2$ is often chosen. \cite{Warren2006} demonstrated that FoF linking would underestimate the masses of haloes with low particle number, and proposed a correction that is sometimes applied to boost the halo masses of FoF identified haloes. Other relations between linking length and overdensity have been proposed in the literature \cite[\eg][]{Courtin2011, More2011}. It should be noted that SO finders can also underestimate the halo mass function at low mass due to particle-resolution issues, see \citealt{Nishimichi2019}.

\subsection{Baryonic feedback}
\label{sec:baryons}

When modelling the matter power spectrum via the halo model it is common to employ NFW profiles, which provide a reasonable match to gravity-only simulated data at small scales. However, in reality `matter' in the universe is comprised of CDM, gas and stars/dust, each of which occupies haloes in a unique way. The halo model can be used to gauge the effect of the presence of gas and stars,  which alter the matter power spectrum compared to the form it would have were gravity to be the only significant force in structure formation. Larger-scale effects ($0.5\iMpc \lesssim k \lesssim 10\iMpc$) on the power spectrum arise from redistributed gas,  mainly due to the Active Galactic Nuclei (AGN) expelling gas from halo centres\footnote{On smaller scales supernovae explosions can also contribute to baryon feedback.}.   Due to these effects, and its intrinsic pressure, gas that remains bound to a halo may have a different profile from NFW.  Smaller-scale contributions ($k \gtrsim 10\iMpc$) to the power deviation arise primarily from stars clustering densely in halo cores \citep[see][for a review of feedback in cosmology]{Chisari2019b}.

Originally, \cite{White2004b} showed that the small-scale matter spectrum could be changed at the $\order{10\%}$ level by reasonable changes to the halo structure that could be calculated theoretically in a spherical-halo scenario with angular-momentum-conserving gas collapse. \cite*{Rudd2008} and \cite*{Zentner2008} show that a decrease in small-scale power was expected due to AGN activity, and that this could be captured by changing the concentration--mass relation in the NFW profile that enters the halo-model power spectrum calculation. The exact impact that feedback has on the matter spectrum is still uncertain \citep[\eg][]{vanDaalen2011, McCarthy2017, vanDaalen2020}, but is certainly at least $\order{10\%}$ for reasonable feedback scenarios. 

\cite{Semboloni2011, Semboloni2013, Fedeli2014a, Fedeli2014b} show that total-matter power spectra can be constructed by taking separate profiles for each component of the matter, and that the impact of AGN feedback could be captured if the gas content of haloes was assumed to be decreased. \cite{Mead2020} showed that this could be extended to modelling all combinations of auto/cross spectra that can be extracted from the hydrodynamic simulations. This modelling can then form the basis of effective models of feedback that attempt to model the response (or reaction) in power spectrum only \citep[\eg][]{Mead2020, Mead2021a}, thus circumventing the difficult issue of the general inaccuracy of the halo-model calculation. 

In approaches such as \cite{Mohammed2014a, Mohammed2014b, Sullivan2021} where the one-halo term is reduced to a series expansion, it has been shown that the series can be fitted to power spectra for a range of feedback scenarios with similar performance to the gravity-only case. \cite{Debackere2020} suggested that the mass-dependent halo baryon fraction could be measured using external data (\eg thermal or kinetic Sunyaev-Zeldovich or X-ray observations), and this could be used to provide an external constraint on the impact that feedback may have on the power. However, such arguments rely on the halo model providing a perfect mapping between the properties of haloes and their power spectra.

How baryonic feedback alters the spectrum of tracers other than matter has not received significant attention.   In most cases, galaxy--galaxy lensing and galaxy clustering studies have limited themselves to large scales where the impact of baryon feedback and non-linear galaxy bias are small.  Although, feedback has been accounted for in studies that push to smaller scales by allowing for a variable halo-concentration amplitude (accounting for matter redistribution) and a separate concentration amplitude for the satellite galaxies \citep[see for example][]{Cacciato2013,Viola2015, vanUitert2016, Dvornik2018, Debackere2020,Dvornik2023,Amon2023}.

\subsection{Modelling the matter power spectrum}
\label{sec:modelling_matter}

Modelling the matter power spectrum is particularly useful for weak-lensing studies, where the lensing signal is sourced by the distribution of \emph{all} matter in the universe. However, it has long been recognised that the accuracy of the halo model prediction is poor compared to what is required by contemporary lensing data (see Fig.~\ref{fig:hm_comp}). This has led to several attempts to develop fitting functions specifically for the matter power.

\subsubsection{\halofit}
\label{sec:halofit}
Originally presented by \cite{Smith2003}, \halofit is a halo-model-inspired fitting function with $\sim 30$ free parameters that was fitted to \nbody simulation data. It does not use the halo model directly, but the power spectrum is broken down as the sum of a `quasi-linear' and `halo' term, which are analogues of the two- and one-halo terms. Further inspiration from the halo model is used in that the fitting functions are parameterised in terms of $\sigma(R)$ (and its derivatives), as opposed to random functions of the cosmological parameters. \halofit was updated in accuracy by \cite{Takahashi2012} and a prescription for massive neutrinos was added by \cite{Bird2012}. \halofit is accurate at around the $\sim5\%$ level for $k<10\iMpc$ and $z<2$ for a wide range of cosmologies. Note well that \halofit cannot be used to predict any spectra other than that of matter.

\subsubsection{\hmcode}
\label{sec:hmcode}
Originally presented by \cite{Mead2015b} and then updated by \cite{Mead2016} and \cite{Mead2021a}, \hmcode is a version of the halo model that has been augmented to produce accurate matter power spectra. While the backbone of the calculation is the vanilla halo model described in Subsection~\ref{sec:matter}, there are several additions and tweaks that were necessary in order to enhance accuracy. These tweaks ensure that the model pertains to a population of `effective haloes' the physical reality of which should not be taken too seriously. \hmcode is accurate at the $\sim2.5\%$ level for $k<10\iMpc$ and $z<2$ across a wide range of cosmologies. Note that \hmcode cannot be used to predict any spectra other than that of matter. It is also not obvious that the same tweaks that are required to provide accurate matter spectra would work, or would even be appropriate, if one wanted to extend the method to other tracers.

\section{Modelling tracers}
\label{sec:tracers}

So far, we have discussed the application of the halo model for calculating the power spectrum of matter and galaxies, but we noted in Section~\ref{sec:derivation} that our initial derivation was applicable to \emph{any} diffuse tracer of large-scale structure whose halo profile can be specified. In this Section we review work where the halo model has been used to calculate these other spectra. Most authors simply replace the halo profiles in equations~(\ref{eq:one_halo_term}) and (\ref{eq:two_halo_term}; ignoring $\Bnl$) with those relevant for the new tracer. There is no need to specify new mass functions or halo-bias relations since, in a model where all signal originates from haloes, this is already included self consistently. Generally, the spectrum of the new tracer will have the linear shape at large scales, with an amplitude determined by the tracer occupation statistics. At smaller scales the shape of the one-halo term will be governed by the shape of the tracer profiles. Using the halo model in this way might be accurate, but the accuracy should be assessed on a case-by-case basis and ideally should be confirmed by comparing the results of calculations to measurements from simulations. Some of the additions that will be discussed in Section~\ref{sec:non_standard} may be more or less important for spectra of different tracers. Finally, we note that any significant contribution to a signal that is genuinely diffuse, such that it cannot be tied to a halo, is difficult to include self consistently within a model where all signal originates from haloes (although see Section~\ref{sec:smooth}).

\subsection{Galaxies}
\label{sec:HOD}

We touched on modelling galaxy power spectra in Section~\ref{sec:discrete_tracers}.  Here we go into more detail and explore the relation between halo masses and the distribution of galaxies within those haloes, the so-called halo occupation distribution (HOD).  Note that, if a stochastic relationship between galaxy occupation and halo mass is assumed then properties of the statistical distribution of galaxies also need to be specified.

A commonly-used HOD is the five-parameter model of \cite{Zheng2005}, who measured the relation between haloes and galaxies from a smoothed particle hydrodynamics simulation and a semi-analytic galaxy formation model.  They find that the mean number of central galaxies given a halo mass, $M$, can be well described by
\begin{equation}
\average{N_\mathrm{c}(M)} = \frac{1}{2}
\left[1+\mathrm{erf}\left(\frac{\log_{10}(M/M_\mathrm{min})}{\sigma_{\log_{10}M}}\right)\right]\ ,
\label{eq:HOD_Nc}
\end{equation}
where $\mathrm{erf}(x)$,  is the error function. 
Haloes can never host more than a single central galaxy, as enforced by the error function ranging from -1 to 1.  
$M_{\rm min}$ is the characteristic minimum halo mass, which means that for $M\ll M_\mathrm{min}$ a halo is unlikely to host a central galaxy while for $M\gg M_\mathrm{min}$ haloes are almost certain to host a single central galaxy, 
with the width of the transition around $M_\mathrm{min}$ governed by $\sigma_{\log_{10}M}$. 
Any random process whose outcome can only be zero or one is governed by Bernoulli statistics; the statistical properties of this distribution are given in Table~\ref{tab:HOD}.

In the  \cite{Zheng2005} model the mean number of satellite galaxies is
\begin{equation}
\average{N_\mathrm{s}(M)} = \Theta(M-M_0)\left(\frac{M-M_0}{M_1}\right)^\alpha\ ,
\label{eq:HOD_Ns}
\end{equation}
where $M_0$ is the truncation mass, below which a halo is not expected to host any satellites ($ \Theta(M-M_0)=1$ for $M>M_  0$ and 0 otherwise).
Haloes with $M=M_0+M_1$ host a single satellite galaxy on average. If $\alpha=1$ then the number of satellite galaxies scales linearly with halo mass. Satellite galaxies are often assumed to follow Poisson statistics\footnote{Although, as we will discuss below, this condition usually does not hold.}; the statistical properties of this distribution are given in Table~\ref{tab:HOD}. A simpler three-parameter HOD is that of \cite{Zehavi2004}, which maps to that of \cite{Zheng2005} in the limit that $\sigma_{\log_{10}M}=0$ in equation~(\ref{eq:HOD_Nc}) and $M_0=0$ in equation~(\ref{eq:HOD_Ns}). A more comprehensive HOD model has been given in \cite{Cacciato2012} which links the distribution of galaxies with their luminosity and stellar mass functions; quantities that can be more readily connected to observations.  Emulator based approaches have also been proposed to model HOD,  allowing for extra free parameters to be included: for example those that capture assembly bias \citep{Salcedo2022}.

\begin{table}
\caption{Halo-occupation properties for central and satellite galaxies assuming that central galaxies are Bernoulli distributed with mean $p$ (equation~\ref{eq:HOD_Nc}) and satellite galaxies are Poisson distributed with mean $\lambda$ (equation~\ref{eq:HOD_Ns}). Also shown is how the statistics of satellite galaxies are modified if the central condition is imposed. In this case the satellite galaxy distribution is no longer Poisson and $\lambda$ is no longer the mean, even if $\lambda$ would usually be obtained from (something similar to) equation~(\ref{eq:HOD_Ns}). If $p=1$ then the central condition has no impact and the expressions in the last two rows of the table are equal. Note also that $\average{N_\mathrm{c}N_\mathrm{s}}=\average{N_\mathrm{s}}=\lambda$, or equivalently $\mathrm{Cov}(N_\mathrm{c}, N_\mathrm{s})=\lambda(1-p)$.}
\begin{center}
\begin{tabular}{c c c c c c}
\hline
Galaxy type & $\average{N}$ & $\average{N^2}$ & $\average{N(N-1)}$ & $\mathrm{Var}(N)$ \\
\hline
Centrals & $p$ & $p$ & $0$ & $p(1-p)$ \\
Satellites & $\lambda$ & $\lambda(1+\lambda)$ & $\lambda^2$ & $\lambda$ \\
cen. cond. & $p\lambda$ & $p\lambda(1+\lambda)$ & $p\lambda^2$ & $p\lambda[1+\lambda(1-p)]$ \\
\hline
\end{tabular}
\end{center}
\label{tab:HOD}
\end{table}

If the parameters $N_\mathrm{c}$ and $N_\mathrm{s}$ were independent then $\average{N_\mathrm{c}N_\mathrm{s}}=\average{N_\mathrm{c}}\average{N_\mathrm{s}}$ in equation~(\ref{eq:one_halo_central_satellite}), but it would be possible for a halo to host a satellite galaxy without hosting a central. To avoid this, the `central condition' is often imposed such that the number of satellite galaxies is fixed to zero if there is no central galaxy.
Imposing this additional constraint affects the (initially assumed) statistics of $N_\mathrm{s}$, especially for halo masses that contain $\sim1$ galaxy, where $N_\mathrm{c}\sim 0.5$. In this case, the central condition distorts the initial distribution,  resulting in fewer haloes containing satellites, and therefore more haloes containing zero satellites than would otherwise be assumed \citep[\eg][]{Beutler2013}. 
This means that the assumption of Bernoulli statistics for central occupation, Poisson statistics for satellite occupation, and the central condition are mutually incompatible. Something has to give, and traditionally it is the Poisson assumption for satellite galaxies that is modified to retain consistency.
If $\mathcal{P}(N_\mathrm{s})$ is the original probability for a halo to host $N_\mathrm{s}$ satellite galaxies, then this is modified according to
\begin{equation}
\mathcal{P}'(N'_\mathrm{s}) = \average{N_\mathrm{c}} \mathcal{P}(N_\mathrm{s})+(1-\average{N_\mathrm{c}})\delta_{0 N_\mathrm{s}}\ ,
\end{equation}
where $\delta_{ij}$ is the Kroenecker delta. If $\average{N_\mathrm{c}}=1$ then the satellite distribution is unchanged; if $\average{N_\mathrm{c}}=0$ then $\mathcal{P}'(N'_\mathrm{s}=0) = 1$ and $\mathcal{P}'(N'_\mathrm{s}\geq 1)=0$ automatically. These transformations of the probability distribution modify $\average{N_\mathrm{s}}$, $\average{N^2_\mathrm{s}}$ and $\average{N_\mathrm{s}(N_\mathrm{s}-1)}$ in a calculable way, with each being lowered by a multiplicative factor of $\average{N_\mathrm{c}}$ relative to the values calculated for the initially assumed distribution. Therefore, if applying the central condition, the replacements in the final row of Table~\ref{tab:HOD} should be made when evaluating equations~(\ref{eq:one_halo_satellite_satellite}) and (\ref{eq:one_halo_central_satellite}). Note that this implies that $\average{N_\mathrm{s}}$ calculated from equation~(\ref{eq:HOD_Ns}; or any similar equation) will \emph{not} be the true mean number of satellites, and that a covariance between $N_\mathrm{c}$ and $N_\mathrm{s}$ is generated. Recent studies also show that the distribution of galaxies can be significantly non-Poissonian in a mass-dependant manner, and if ignored this can potentially bias cosmological results \citep{Dvornik2018, Avila2020,Beltz-Mohrmann2020,Hadzhiyska2022, Dvornik2023}.

\begin{figure}
\begin{center}
\includegraphics[width=\columnwidth]{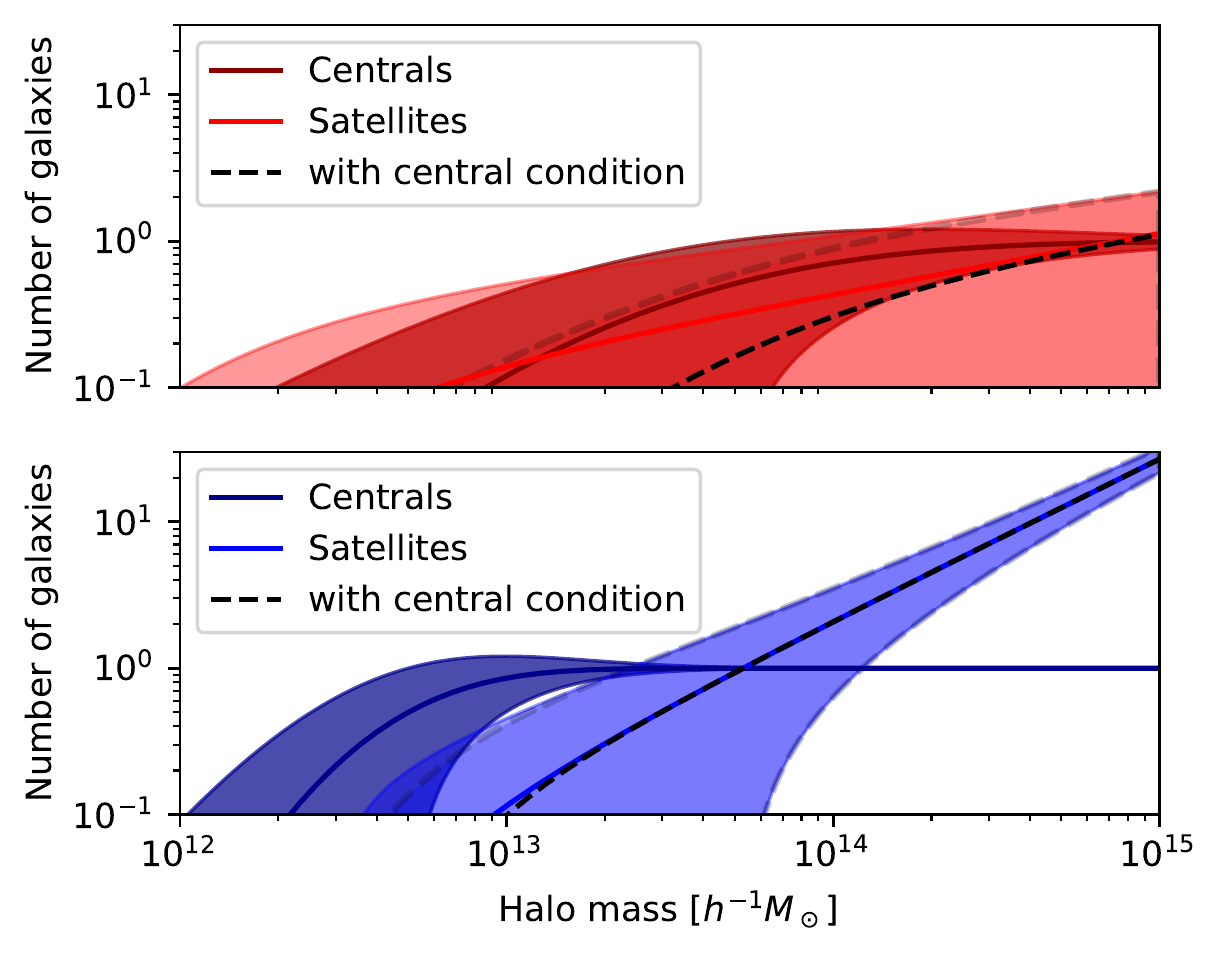}
\end{center}
\caption{Example HODs for typical surveys that targets red galaxies (upper) and blue galaxies (lower). The solid lines show the mean numbers of central and satellite galaxies as a function of halo mass, while the bounded regions show the expected scatter ($\pm\sigma$) about this.  Note that the shaded region for central occupation includes values beyond 1,  although this is a plotting artefact.
The darker dashed line shows the number of satellite galaxies when imposing the central condition while the lighter dashed lines indicate their expected scatter.  The number of satellites is generally lowered when this condition is imposed.  This is a small effect, and it induces a covariance between central and satellite galaxies in the region where $\average{N_\mathrm{c}}\sim0.5$ and the scatter is also affected. When $\average{N_\mathrm{c}}\sim1$ this difference vanishes.  }
\label{fig:HOD}
\end{figure}

Example HODs are shown in Fig.~\ref{fig:HOD} where the small difference to the mean and variance of the satellite galaxy distribution when imposing the central condition can be seen. We show example HODs from surveys that target red (\eg BOSS; parameters taken from \citealt{Zhai2017}) and mostly blue (\eg GAMA; parameters taken from \citealt{Smith2017}) galaxies. Red galaxies are mainly centrals, with a few satellites, while blue galaxies are mainly satellites at high halo mass\footnote{The parameters used for the mostly blue sample are taken from Figure 4 of \cite{Smith2017} for r-band magnitude $-21$.  }.  There are more suitable HOD models that describe purely blue samples,  such as star forming galaxies which are normally identified as emission line galaxies \citep[for example][]{Avila2020}.

\subsection{Intrinsic Alignments of Galaxies}
Galaxy formation processes are expected to imprint a correlation between the ellipticity of a galaxy and its environment, hence also with its neighbours.  As the observed correlation between galaxy shapes is the core measurement to detect weak gravitational lensing by large-scale structure, it is critical to determine the correlations that arise purely from the intrinsic alignment (IA) of galaxies.  Without accurate IA models, robust cosmological constraints can not be extracted from weak lensing observations \citep[see][for a review]{joachimi/etal:2015}.  

Luminous red galaxies have been observed to align with their local density field \citep[see for example][]{mandelbaum/etal:2006,joachimi/etal:2011}, but the intrinsic alignment between blue galaxies has yet to be detected \citep{johnston/etal:2019}.  This suggests that different shape-formation mechanisms may be in place.  One hypothesis is that the ellipticity of red/elliptical galaxies is determined by the linear tidal field \citep{catelan/etal:2001, hirata/seljak:2004}, with blue/spiral galaxies gaining their shape through a tidal-torque mechanism \citep{schafer:2009}.  \citet{Schneider2010} suggest that satellite galaxies may also be subject to an infall alignment mechanism where their ellipticity points towards the centre of their parent halo.  Observations find a more complex scenario with satellite radial alignment detected on small scales, changing to a random alignment on large scales \citep{georgiou/etal:2019}, with the alignment strength also sensitive to luminosity \citep{singh/etal:2015,huang/etal:2018}.

The halo model framework provides a route to encode this complexity to define a flexible IA cosmological model \citep{Schneider2010}.  
\citet{Fortuna2021} adopt the commonly-used \citet{bridle/king:2007} Non-linear Linear Alignment model (NLA) to describe the alignment between central galaxies.  The NLA model is based on the \citet{hirata/seljak:2004} linear tidal field alignment model where galaxy shape is determined not from the properties of its parent halo, but from the external density field.  This approach allows us to bypass the need to model halo asphericity (see Section~\ref{sec:asphericity}) that is known to correlate with galaxy shape in hydrodynamical simulations \citep{chisari/etal:2017,xi/etal:2023}.  The NLA model augments the linear model by replacing the linear matter power spectrum with a non-linear version, such as \hmcode, which was found to improve the agreement of the model with IA numerical simulations \citep{heymans/etal:2006}. The NLA matter-intrinsic ellipticity power spectrum, commonly referred to as `GI',  is given by 
\begin{equation}
P^{\rm NLA}_{\rm GI}(k) = -A_{\rm IA}\left(\frac{1+z}{1+z_{\rm pivot}}\right)^{\eta} \frac{C_1\rho_{\rm crit}\Omega_{\rm m}}{D} P^{\rm non-linear}_{\rm mm} (k)\;,
\label{eq: LA}
\end{equation}
where the amplitude, $A_{\rm IA}$, and the redshift evolution parameter, $\eta$, are free parameters that are usually fitted to the data,  $D$ is the linear growth factor, $z_{\rm pivot}$ is an arbitrary pivot redshift and $C_1$ is a constant defined to match early IA observations \citep{brown/etal:2002}.
\citet{Fortuna2021} then use an HOD to estimate the fraction of blue and red centrals to determine the `two-halo' part of the matter-intrinsic ellipticity power spectrum with
\begin{equation}
P^{\rm 2h}_{\rm GI} (k) = f^{\rm red}_{\rm cen} P^{\rm NLA, red}_{\rm GI}(k)  + f^{\rm blue}_{\rm cen} P^{\rm NLA, blue}_{\rm GI}(k) \, .
\end{equation}
Here different amplitudes (and sometimes $\eta$) for the NLA model of the red and blue central populations are facilitated, although \citet{Fortuna2021} show that this may not be necessary as the colour dependence of the IA signal is not seen when restricting the sample to central galaxies.  

As the halo population is, on average, spherical, the average inter-halo satellite-central alignment is zero. The `one-halo' IA term then derives from the alignment of satellites with each other and the local matter field with
\begin{equation}
P^{\rm 1h}_{\rm GI} (k) = \int\, \mathrm{d}M \, n(M) \frac{M}{\bar{\rho}_{\rm m}} f_{\rm s} \frac{\langle N_{\rm s}|M \rangle}{\bar{n}_{\rm g}}\, |\hat{\gamma}^{\rm I}({\bf k}|M)| \, \hat{U}(M,k) \, .
\end{equation}
Here $n(M)$ is the halo mass function (Section~\ref{sec:halo_mass_function}), $f_{\rm s}$ is the fraction of satellites which may vary as a function of redshift, $\langle N_{\rm s}|M \rangle$ is the halo occupation distribution of satellites (equation~\ref{eq:HOD_Ns}), ${\bar{n}_{\rm g}}$ is the mean number density of galaxies (equation~\ref{eqn:ng}), and $\hat{U}(M,k)$ is the Fourier transform of the normalised matter density profile (equation~\ref{eq:normU}).  The alignment strength depends on $\hat{\gamma}^{\rm I}$, the density weighted average of the projected satellite ellipticity, assuming all satellites point towards the halo centre.  This term can also include a radial dependence, to decorrelate the alignment on large scales.

In this section we have outlined the halo model for the correlation between intrinsic galaxy ellipticity and the density field, the `GI' term.  We refer the reader to \citet{Schneider2010} for the equivalent terms for the correlation between intrinsic shapes, also known as the `II' term.  Both terms contaminate a tomographic weak-lensing analysis.  \citet{Fortuna2021} argues that this halo model approach is preferable to the most often used NLA model, as it provides a natural route to include observations of the changing fraction of red and blue galaxies across the redshift range of the weak lensing survey, in addition to direct measurements of satellite alignments within groups.  
  
\subsection{Thermal Sunyaev-Zeldovich effect}
\label{sec:tSZ}

The thermal Sunyaev-Zeldovich (tSZ) effect arises when CMB photons are scattered by free electrons, predominantly those hot electrons found in galaxy clusters. This results in a spectral distortion of the CMB black-body spectrum with a characteristic frequency dependence, and this (Compton-$y$ signal) can be extracted from CMB temperature data. The strength of the $y$ signal depends on the product of the free electron temperature and density, a quantity that has units of pressure, and it is therefore the electron-pressure profile that is relevant for tSZ halo-model calculations. The number of free electrons in a halo scales with $M$, and the temperature scales as $\sim M^{2/3}$, so the overall profiles scales like $\sim M^{5/3}$, which means that the shape and amplitude of spectra involving $tSZ$ are determined by more massive haloes than either matter or galaxies, whose profiles scale exactly as $M$ and $\sim M$ (if satellite dominated) respectively. This in turn implies that spectra involving tSZ are relatively sensitive to $\sigma_8$ \citep{Refregier2002, Komatsu2002}, which arises because the high-mass end of the halo mass function is sensitive to $\sigma_8$. This also means that the one-halo term is comparatively high amplitude, and the transition region in the power spectrum occurs at a relatively larger scale \citep[\eg][]{Mead2020}. The dependence on gas temperature makes tSZ an interesting direct probe of baryonic feedback \citep[\eg][]{McCarthy2014, Hojjati2015}. A good pedagogical discussion of the halo model in the context of tSZ is provided by \cite{Hill2013b}, as well as the idea of masking massive low-redshift clusters in order to boost the signal-to-noise (see also \citealt{Hill2018}). 

Electron pressure profiles can be derived from theoretical arguments \citep[\eg][]{Komatsu2001, Ostriker2005} or from fitting to observational data and simulations \citep[\eg][the so-called \emph{universal} pressure profile]{Arnaud2010}. It is clear that concepts like non-thermal pressure support and baryonic feedback affect the pressure distribution with galaxy clusters \citep[\eg][]{Shaw2010}, and therefore that models based on hydrostatic equilibrium are overly simplified.

While the tSZ auto spectrum can be measured, the cosmological constraints from this are in disagreement with those from more developed probes \citep[\eg][]{Planck2015XXII}, possibly due to the self-correlation of residual systematics in the $y$ maps (although see \citealt{McCarthy2014, Horowitz2017, Bolliet2018}). With the auto spectrum suspect, tSZ halo models have been used in cross correlation by: \citeauthor{Addison2012} (\citeyear{Addison2012}; Cosmic Infrared Background CIB);  \citeauthor{Hajian2013} (\citeyear{Hajian2013}; $X$-ray clusters); \citeauthor{Hill2014} (\citeyear{Hill2014}; CMB lensing); \citeauthor{Ma2015} (\citeyear{Ma2015}; galaxy lensing); \citeauthor{Vikram2016} (\citeyear{Vikram2016}; galaxy groups); \citeauthor{Tanimura2019b} (\citeyear{Tanimura2019b}; galaxy clustering); \citeauthor{Koukoufilippas2020} (\citeyear{Koukoufilippas2020}; galaxy clustering); \citeauthor{Yan2021} (\citeyear{Yan2021}; CIB, galaxy clustering); \citeauthor{Maniyar2021} (\citeyear{Maniyar2021}; CIB). Much of the focus is on measuring the pressure bias from the large-scale portion of the power spectrum, which can be thought of as the $k\to0$ limit of the pressure integral that contributes to the two-halo term
\begin{equation}
\average{bP_\mathrm{e}} = \int_0^\infty P_\mathrm{e}(M)b(M) n(M)\,\diff M\ ,
\end{equation}
where $P_\mathrm{e}(M)$ is the mean electron pressure in a halo of mass $M$. Other focus is on the so-called \emph{hydrostatic-mass bias}, which arises due to deviations of the gas from hydrostatic equilibrium, and this biases inferred cluster masses compared to more direct (\eg weak lensing) measurements.

\subsection{$X$-rays}
\label{sec:X-ray}

$X$-rays are emitted via the bremsstrahlung process, when the direction-of-travel of free electrons is modified by an interaction with a proton. Since this is a two-body scattering process the total contribution of a halo profile scales like $ M^2\times T \sim M^{8/3}$, even more strongly than electron pressure. This ensures that the $X$-ray auto spectrum will be completely dominated by the one-halo contribution (unless masking is applied), which in turn leads to a strong dependence on $\sigma_8$ \citep{Diego2003} via the dependence of the high-mass tail of the halo mass function. Halo models of $X$-rays maps have been used in cross-correlation analyses by: \citeauthor{Hurier2015} (\citeyear{Hurier2014}; \citeyear{Hurier2015}; tSZ); \citeauthor{Singh2017} (\citeyear{Singh2017}; AGN galaxy clustering); \citeauthor{Hurier2019} (\citeyear{Hurier2019}; CMB lensing).

\subsection{Cosmic Infrared Background}
\label{sec:CIB}

Warm dust in star-forming galaxies emits infrared radiation, which can be detected by space-based telescopes. The halo signal will be proportional to the amount of dust in a halo, which in simple models can be taken to scale with the number of galaxies \citep[\eg][]{Xia2012}, although different populations of galaxies (\eg spirals, star-burst, proto-spheroids) can also be considered as long as an occupation model for each is specified. Given the galactic origin, one may expect the halo profile of CIB emission to be similar to that of galaxies. Note that CIB flux is typically measured in frequency bins, and CIB emission from distant galaxies will be redshifted, so the resulting angular power spectra will mix galaxy populations and redshifts in unintuitive ways. \cite{Xia2012} considered the CIB power spectrum while \cite*{Addison2012} considered the cross spectrum of CIB and tSZ. \cite*{Addison2013} suggested that evolution of the dust spectral energy distribution and scale-dependent halo bias (\ie $\Bnl$) may be required to self consistently understand one- and two-point functions of the CIB within the same theoretical framework. More recently, \cite{Maniyar2021} has presented a halo model where CIB emission is tied to the halo-mass accretion history.

\subsection{Neutral hydrogen}
\label{sec:HI}

Neutral hydrogen (\textsc{hi}) emits characteristic $21\cm$ wavelength radiation when the electron and proton spin align or misalign. The profile signal will scale with $M$ and the fraction of \textsc{hi} in a halo, but \textsc{hi} is eroded by heat and AGN activity, so the signal will be dominated by comparatively low-mass haloes that still retain significant \textsc{hi} reservoirs. Ingredient lists for the \textsc{hi} abundance and halo profiles can be found in \cite{Padmanabhan2017a} and \cite{Villaescusa-Navarro2018}, and clustering calculations are presented in \cite{Padmanabhan2017b}, \citeauthor{Feng2017} (\citeyear{Feng2017}; in cross-correlation with Lyman-$\alpha$) and \cite*{Schneider2021}. \cite{Wolz2019} investigates shot noise in the power spectrum of \textsc{hi} under differing assumptions about the source of \textsc{hi} emission -- either co-located with galaxies or with dark matter.


\subsection{$\gamma$-rays}
\label{sec:gamma}

If dark matter has a significant self-interaction cross section then dark matter--dark matter annihilation events are expected to produce a potentially detectable flux of $\gamma$-rays. This signal scales with the square of the dark-matter density, with the result that halo cores, low-mass haloes and subhaloes are expected to produce the most significant contributions. The cross correlation of gamma ray maps with large-scale structure has been investigated by \cite{Shirasaki2014} and \cite{Troester2017} using halo models to generate theory curves and constrain the cross section.

\section{Extensions to the standard halo model}
\label{sec:non_standard}

\begin{figure}
\begin{center}
\includegraphics[width=\columnwidth]{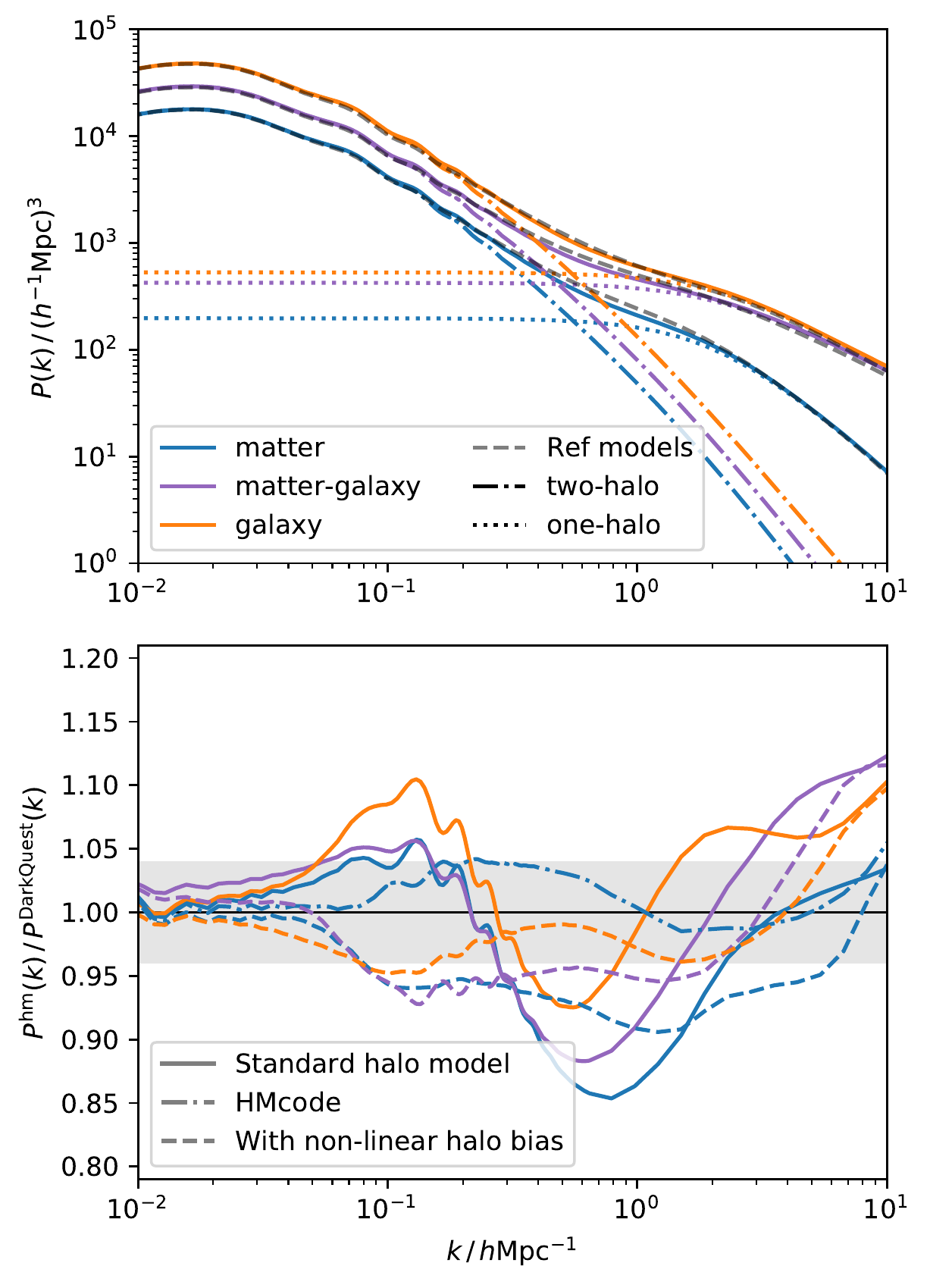}
\end{center}
\caption{The performance of the analytical halo model calculation compared to state-of-the-art semi-analytical calculations from \darkemu at $z=0$ for matter, galaxy and matter-galaxy power spectra.  \darkemu is accurate at the $4\%$ level relative to simulations (shaded region).  The upper panel shows the total $P(k)$ (solid) and the one-halo (dotted) and two-halo (dot-dashed) terms.  The lower panel shows the ratio between the halo model and the \darkemu power spectra. The vanilla halo model predictions (solid curves) show errors of between  $\sim10\%$ and  $\sim15\%$ with respect to the reference model.  When the non-linear halo bias,  calculated using the same emulator, is included (dashed lines) the error is reduced to $<10\%$ for intermediate scales.  Including this effect in the halo model works best for the galaxy power spectrum bringing its  error to within  $\sim5\%$ of the reference model.  This is because the dark emulator can only estimate the halo-halo power spectrum for more massive haloes, which host most of the galaxies that contribute to the signal.  When modelling matter or matter-galaxy power spectra,  an estimate of the clustering of lower mass haloes is also needed.  As a result we fail to achieve the same level of accuracy for these power spectra.  The \hmcode prediction on the other hand provides a good agreement with the \darkemu matter power spectrum.  \hmcode is accurate at the $2.5\%$ level compared to the simulations it was calibrated against.}
\label{fig:hm_comp}
\end{figure}

The standard derivation outlined in Section~\ref{sec:derivation}, its extension to discrete tracers outlined in Section \ref{sec:discrete_tracers} and sets of ingredients like those mentioned in Section~\ref{sec:ingredients}, constitutes most of what is commonly called `the halo model' in the literature. In Fig.~\ref{fig:hm_comp} we demonstrate the performance of the standard calculation (\LCDM; $z=0$; $\Delta_\mathrm{h}=200$; \citealt{Tinker2010} mass function and bias; \citealt{Duffy2008} concentration--mass relation; \citealt{Zheng2005} satellite-dominated HOD with $M_\mathrm{min}=10^{13}\Msun$; $\sigma=0.5$; $M_0=M_\mathrm{min}$; $M_1=10^{14}\Msun$; $\alpha=1$) compared to state-of-the-art semi-analytical theoretical calculations for matter, galaxy and matter-galaxy power from the \darkemu \citep{Nishimichi2019}, which is accurate at the $\sim4\%$ level relative to simulations.  We see that the analytical calculation is in error at the $\sim15\%$ level, with a similarly-shaped residual for each power spectra. 
At $k\sim0.03\iMpc$ we see good agreement, where differences between the halo model and \darkemu originate from differences in linear halo bias.  At $k\sim0.1\iMpc$ we see the standard halo models \emph{over} predict the power, which is due to perturbative effects suppressing the true power at these scales, but this not being included in the halo model. Including the non-linear halo bias ameliorates this because the non-linear halo clustering contains these non-linear effects. At $k\sim0.7\iMpc$ all halo models under-predict the power, which is somewhat corrected once the non-linear halo bias is added.  For $k>2\iMpc$ the match is quite good, but degrades at smaller scales where the shapes of the halo profiles in the one-halo term become important. For the matter power we also compare  \hmcode \citep{Mead2021a} with  \darkemu and find them in good agreement.  \hmcode is accurate at the $\sim2.5\%$ level relative to simulations. 

To improve these discrepancies, many authors have considered extensions to the standard model, and we discuss some of these here. These extensions will usually require some new ingredient to be calibrated from simulations. The following subsections here are loosely ordered in terms of the scale they have an impact on, from largest to smallest. Some of the non-standard approaches only apply to power spectra for a specific tracer combination, while others offer more general potential improvements.

\subsection{Smooth matter}
\label{sec:smooth}

When a halo finder is run on \nbody simulation output, it commonly allocates $\sim$half of the simulated mass to haloes. The remaining half is either in structures of low particle number (a threshold of $\sim200$ particles is often adopted) or else looks to be approximately smoothly distributed in the inter-halo medium. If the simulation is re-run with more \nbody particles low-mass haloes are often identified in regions that were previously thought to be devoid of haloes. It remains an unsolved problem in cosmology to determine whether all mass in the Universe is bound in haloes of ever lower masses, although there are compelling theoretical arguments \citep[\eg][]{Press1974, Bond1991}, and simulations suggest that this should be so \citep[\eg][]{Angulo2017, Wang2020}, at least for perfectly cold dark matter particles\footnote{ Although note that any realistic particle-based dark-matter model has at least some streaming velocities, which translates to some unbound particles}.

Halo mass functions are fitted to data from \emph{measured} haloes, above some mass threshold, and sometimes the fitting forms are constrained such that all mass in the universe \emph{is} contained in haloes if the fitting function is interpreted literally \citep[\eg][]{Sheth1999}. Some authors have considered how a genuinely smooth component of matter could be added to the halo model, either in the context of WDM \citep{Smith2011b}, or in the context of gas expelled from haloes by AGN feedback \citep{Fedeli2014a, Debackere2020, Mead2020}. In all cases the smooth matter is taken to be distributed as per the linear matter perturbations, via a biased linear power spectrum. In the case of matter spectra, the bias of the smooth component is constrained by the fact that an unbiased linear spectrum must be returned at large scales.

\cite{vanDaalen2015} used \nbody experiments to test the contribution of smooth (or non-halo) matter to the overall matter power spectrum. If only matter in FoF haloes (sometimes called groups) is used, then essentially all power for $k>3\iMpc$ can be accounted for, but if SO haloes are used then this drops to between $80$ and $95\%$ of the power, depending on the overdensity threshold. In the SO case, presumably the remaining power arises from dense material just outside the (artificial) spherical halo boundary, which is contained within FoF groups. It is less clear how this extra-halo material contributes at intermediate scales, or if is better to think of it as genuinely smooth matter, or matter that would resolve itself into low-mass haloes were the simulation to be run with higher resolution.

\subsection{Voids}
\label{sec:voids}

In the canonical (linear halo bias) halo model, only `linear' voids will be present, which enter through the lack of  haloes in low-density regions. However, it is clear from visual inspection of the particle distribution in \nbody simulations that the distribution of voids in the Universe is not a Gaussian random field in detail. \cite*{Voivodic2020} create a version of the halo model that explicitly accounts for voids and which therefore requires a void abundance, linear bias and profile to be specified. Voids are essentially considered in the same way as haloes, but with under-density profiles rather than over-density profiles. This then allows for the calculation of correlations such as halo--void, void--void etc. In addition, `dust' (non-halo, non-void) matter is also accounted for in the same manner as described in Subsection~\ref{sec:smooth}. Because voids are physically larger than haloes, their `one-halo' (one-void) term peaks at larger scales than the standard one-halo term, and therefore adds some (much-needed) power into the transition region of the matter power spectrum. Note that in the non-linear halo bias approach advocated in Section~\ref{sec:non_linear_bias}, the two-point contribution voids will be included in the $\beta^\mathrm{nl}$ function, since voids are really a spatially-coherent lack of haloes, and it is the two-point function of this spatial coherence that is captured by $\beta^\mathrm{nl}$.

\subsection{Perturbation theory}
\label{sec:perturbation}

To go beyond linear halo bias and linear power \cite{Valageas2011, Mohammed2014a, Seljak2015, Schmidt2016, Hand2017, Philcox2020} have all considered replacing the linear theory spectrum with something higher-order that can, in principle, be calculated via perturbation theory. These approaches have been demonstrated to improve the halo-model prediction for the matter spectrum in the quasi-linear regime, but it is not obvious how to translate their results to power spectra of tracers other than the matter -- if linear bias is employed the problem indicated in Fig.~\ref{fig:Rhh} remains. \cite{Smith2007} jointly consider a perturbative bias expansion with perturbation theory for the matter, getting good results in the mildly non-linear regime. Recently, \cite*{Sullivan2021} have looked at biased tracers together with a perturbative model at large scales, with an exclusion term incorporated for halo and galaxy clustering. However, the bias model is still effectively linear and fitting parameters mask the underlying physical relationship between halo, matter, and galaxy clustering.

\cite{Valageas2011} combined perturbation theory with the halo model to improve accuracy on quasi-linear scales.
Other schemes of which we are aware that are halo-model inspired and that can be used for matter spectra are the halo-Zel'dovich perturbation theory of \cite{Mohammed2014a}, \cite{Seljak2015}, and \cite{Sullivan2021}. Here the two-halo term is modelled using the \cite{Zeldovich1970} approximation, while a series expansion is used for the one-halo term with halo-model insight used to constrain the allowed terms in the series. Finally, there is the effective halo model of \cite{Philcox2020}, which uses the effective field theory of large-scale structure for a two-halo term combined with a standard one-halo term.

\subsection{Excess large-scale power}
\label{sec:excess_large_scale_power}

On large scales, the Fourier halo profiles $\hat{W}_{u}(M,k\to0) \propto k^0$ (constant) and therefore the same is true of the scale-dependence of the one-halo term (equation~\ref{eq:one_halo_term}): $P_{uv}^\mathrm{1h}(k\to0) \propto k^0$, a shot-noise-like contribution. In contrast, in the same limit the two-halo term (equation~\ref{eq:two_halo_term}) always has the linear spectrum shape at large scales: $P_{uv}^\mathrm{2h}(k\to0) \propto k^{n_s}$. It is then inevitable that the one-halo term has more power than the two-halo term at some sufficiently large scale, with the exact scale being governed by the redshift and the fields $u$ and $v$. This problem manifests differently in configuration space, since terms $\propto k^0$ contribute only to the $r=0$ part of the correlation function. However, this can cause the integral over the correlation function to be non-zero when $r=0$ is included, contrary to the requirement that the mean overdensity be zero.

The standard halo model thus predicts that the total power spectrum transitions from being dominated by the \emph{one-halo} term at ultra-large scales (typically $k\sim10^{-4}\iMpc$), to two-halo power at large scales ($k\sim10^{-2}$) and back to one-halo power at small scales ($k\sim1\iMpc$). For the autospectrum of a discrete tracer this might be the correct manifestation of shot noise dominating the signal at very large scales \citep[see for example][]{Seljak2009}.  For matter--matter, or matter--galaxies this is wrong \citep[\eg][]{Ginzburg2017}, since we would expect intra-halo power to dominate only at small scales, leaving only inter-halo power at large scales\footnote{In practice this rarely affects halo-model predictions since the scales on which this problem is manifest are so large.}. This problem can be seen at the largest scales shown in Fig.~\ref{fig:power}.  Some authors have suggested corrections to halo exclusion (see section~\ref{sec:exclusion},  \citealt{Smith2011a, Schneider2023}) or the interpretation of the 1-halo term \citep{Valageas2011} as possible solutions to this problem. 

\cite{Zeldovich1970} noted that any local process that satisfies mass and momentum conservation must have a large-scale spectrum that decreases faster than $k^4$. Galaxies do not obey these conservation laws, so can have a shot-noise contribution at large scales. However, matter should obey these conservation laws, but they have not been imposed in the standard formulation of the halo model, thus leading to the large-scale breakage \citep{Seljak2000}. Indeed, any shot-noise contribution to the matter one-halo term at large scales is \emph{not} observed in simulations \citep{Crocce2008}. In the effective models of \cite{Mohammed2014a, Seljak2015, Hand2017, Sullivan2021} an ad-hoc `compensation kernel' multiplies the standard one-halo term to enforce mass conservation, $k^2$, and to ensure that predictions from perturbation theory are recovered on large scales. $k^2$ scaling is sufficient to make the one-halo term subdominant at large scales, and it was found that enforcing momentum conservation spoiled predictions in the transition region. Presumably $k^4$ scaling could be enforced with a different form of the compensation kernel if necessary.

\cite{Valageas2011} demonstrated that mass conservation could be imposed in the halo model if a Lagrangian framework was adopted from the outset, resulting in a one-halo term that decays like $k^2$ at large scales, while \cite{Schmidt2016} enforced mass and momentum conservation and was able to get $k^4$ scaling. \cite{Hamaus2010} showed that the large-scale power of massive haloes is sub-Poisson. \cite{Ginzburg2017} was able to incorporate this within the halo model using a halo-fluctuation field and the general bias expansion.

\subsection{Halo compensation}
\label{sec:compensation}

\cite{Cooray2002} suggest using `compensated' matter profiles as a way of avoiding excess large-scale power in matter spectra. These profiles are compensated in the sense that they have both a positive and a negative overdensity that exactly cancel, in contrast to standard profiles that only have positive overdensity. For a compensated matter profile $\hat{W}_{m}(M,k\to0) \propto k^2$ (rather than constant) and therefore $P_\mathrm{mm}^\mathrm{1h}(k\to0) \propto k^4$, in line with the expectation from mass and momentum conservation \citep[\eg][]{Smith2003}. This occurs naturally with compensated profiles, since each is massless and therefore placing and moving these perturbations incurs no mass or momentum cost. However, the compensation also affects the halo-profile that appears in the two-halo term, which therefore also removes the large-scale, linear power from the halo model: $P_\mathrm{mm}^\mathrm{2h}(k\to0) \propto k^{4+n_s}$ rather than $k^{n_s}$; the baby is thrown out with the bath water. This can be understood since compensated profiles have no overall density perturbation when smoothed on scales much larger than their radii. The two-halo term then just re-arranges these zero-density perturbation features, and when smoothed on a sufficiently large scale this will represent an unperturbed density field. A simple fix for this was suggested by \cite{Chen2020, Chen2022}, who view the density field as a sum of the linear perturbations and the haloes\footnote{In the standard halo model the haloes themselves \emph{are} the linear density field.}, if these haloes are compensated then this two-halo term returns to linear form at large scales.

\subsection{Halo exclusion}
\label{sec:exclusion}

One non-linear feature of the halo spectrum that has received considerable attention in the literature is that of `halo exclusion'  \citep[\eg][]{Takada2003, Zehavi2004, Tinker2005, Smith2007, Schneider2012, vandenBosch2013, Baldauf2013}: the fact that haloes have finite radii means that the halo correlation must drop to zero on scales below the radii of the two haloes in question. The details of how this operates depends on the exact definition of haloes \citep[\eg][]{Garcia2019}.  For example, different algorithms may choose to discard halo centres if they are too close.  The eventual masses of haloes may include double counted particles that lie within the boundary of two proximate haloes. 

Exclusion is automatically included if using $P_\mathrm{hh}$ or $\Bnl$ calibrated from simulations \citep[\eg][]{Nishimichi2019, Mead2021b}, but must be incorporated by hand if using a simpler prescription for the two-halo term (\eg linear halo bias).  The scale at which halo exclusion becomes relevant depends on halo size and, by proxy, halo mass.  It will kick in at larger scales for more massive haloes compared to those less massive, and consequently smaller. Since exclusion affects the two-halo term at small scales, it is a subdominant contribution to the total power spectrum for some tracer combinations (\eg matter--matter), but is important for others (\eg galaxy--halo).

Halo exclusion can be approximately included in Fourier space by multiplying the halo power spectrum by some suppressing window function \citep[\eg][]{Schneider2013}. However, most often halo exclusion is included by fixing the halo correlation function to zero below some radius. This strongly suppresses the two-halo term on scales around the halo radius, but ignores the fact that one might expect a more gentle truncation for scales near the halo boundary.  

To allow for a smoother transition, \cite{Zehavi2004} and \cite{Tinker2005} proposed to alter the integration upper limits in equation~(\ref{eq:two_halo_term}) to be dependent on the halo virial radii and the scale in question, in such a way that the halo-halo correlation is fixed to zero if it arises from haloes within some distance of each other: if $r< r_{\mathrm{h},1}+r_{\mathrm{h},2}$. A similar scheme was followed by \cite{Smith2011a}, and this could be expanded to any other function of the halo virial radii. Unfortunately, this makes the halo-model prediction for the overall number density of galaxies deviate from the true value, so the end result needs to be corrected by hand to account for this. In addition, this constraint must be imposed in configuration space, which can make it numerically inefficient if working in Fourier space \citep[][]{Murray2020}.

\subsection{Halo asphericity and substructure}
\label{sec:asphericity}

The standard halo model takes haloes to be perfect spheres with a hard boundary. However, the reality is that haloes have no clear boundary and are generally triaxial (aspherical) with a typical minor-to-major axis ratio of $\sim0.6$ \citep{Jing2002}, furthermore this halo asphericity may be spatially correlated due to gravitational tidal fields. This also applies to galaxies (or other fields) that may exist within haloes, with the galaxy--galaxy and galaxy--matter intrinsic alignments being a particularly important contaminant in weak gravitational lensing \citep[\eg][]{hirata/seljak:2004}. The lowest-order clustering statistics average over all directions, and so a spherical approximation for the halo profile may be reasonable, but recall that $\average{W_u W_v}$ appears in the one-halo term, and therefore any scatter (or covariance) about the mean profile will contribute power. Particular configurations of higher-order statistics are likely to be more sensitive to profile shape than two-point statistics \citep[\eg][]{Takada2003}.

\cite{Smith2005} were the first to lay the theoretical foundations for including asphericty within the halo model. A triaxial profile is required, with distribution functions for the axis ratios, these profiles are then integrated over all orientations in the one- and two-halo terms. An alignment correlation function is required to capture spatial alignments and appears in the two-halo term. In \cite{Smith2005}, halo-model matter power spectra were found to be lower at the $\sim5\%$ level for $k \sim 1$ -- $10\iMpc$ relative to a model with no triaxiality, with the difference mainly driven by high-mass haloes. An alternative approach to environmental dependence is presented by \cite*{Gil-Marin2011} where galaxies are split by environment and which therefore allows cross-correlations between different populations of galaxies to be computed. Tangentially related to these methods are papers that use the halo model to estimate the intrinsic alignment signal \citep[\eg][]{Schneider2010, Fortuna2021} for weak lensing. In these cases, inter-halo intrinsic alignment is driven by large-scale tidal alignment, whereas intra-halo tides align halo satellite galaxies with the central galaxy.

Realistic haloes contain clumpy substructure, but this is ignored when a smooth halo profile is adopted for use in a halo-model calculation. Substructure can be included by breaking the halo into smooth and clumpy components and including a substructure mass function \citep[\eg][]{Sheth2003, Dolney2004, Giocoli2010}. Halo models with substructure generally predict a strong increase in power at small scales ($k\gtrsim10\iMpc$) relative to a model with a smooth halo. Additionally, matter can be broken down into that in clumps and that in the smooth halo, and it is then possible to calculate cross correlations between the clumps (or galaxies therein) and the surrounding matter, which may be more appropriate for galaxy--galaxy lensing.

The impact of halo alignment can also be investigated using simulations: Halo particles (or galaxies) can be randomly rotated about the halo centre to artificially `spherize' the halo; alternatively, individual haloes can be coherently, but randomly, rotated to break intrinsic correlations in the cosmic web. Such numerical experiments were performed by \cite*{vanDaalen2012} for the case of galaxy clustering, where breaking intrinsic alignments was shown to decrease the power spectrum by $\sim2\%$ at $0.1\lesssim k/\iMpc \lesssim1$ and breaking alignments within individual haloes was shown to decrease the power, starting with a $1\%$ level effect at $k\gtrsim0.2\iMpc$ and reaching a more dramatic $\sim20\%$ for $k\simeq10\iMpc$. These are quite significant effects on the power, but clearly the overall impact of these effects will depend on the galaxy sample. To investigate the impact of substructure on the matter power spectrum \cite{Pace2015} performed a similar experiments to \cite{vanDaalen2012}, but for matter profiles, showing that power at $k\gtrsim0.3\iMpc$ was subdued at the $\gtrsim 0.5\%$ level for coherent rotation and $\gtrsim 1\%$ level for spherizing rotation (which removes halo substructure). It was also demonstrated that the two types of rotations have similar ($\gtrsim 1\%$ for $k\gtrsim 0.3\iMpc$) level effects on the cross spectrum between those particles inside and outside of haloes (a proxy for the substructure--matter cross spectrum). These differences increase at smaller scales and are likely to be strongly dependent on the combination of tracers considered. Note well that halo profiles measured from simulations are often subject to the `spherizing' technique discussed in this paragraph if they are measured from the stacked profiles.

\subsection{Scatter in halo properties and assembly bias}
\label{sec:scatter}

When matter profiles are measured from simulations a significant scatter is observed in the relation between halo concentration and mass. For concentration measured for spherical NFW haloes, this scatter is approximately log-normal with $\sigma_{\ln c}\sim 0.3$ \citep[\eg][]{Jing2000, Bullock2001}. The scatter may contain cosmological information if it relates to the scatter in halo-formation times, which itself depends on the power spectrum and can be estimated using extended-Press-Schechter theory \citep[\eg][]{Bond1991}. 

However, this scatter is often ignored in the halo model, and  instead the mean halo profile and the mean concentration--mass relation are used. This is not correct in detail because it is the expectation value of the squared halo profile that appears in the one-halo term (equation~\ref{eq:one_halo_term}). In the limit that this scatter is not spatially correlated, its impact may be accounted for by introducing a second integral within equation~(\ref{eq:one_halo_term}) over the probability distribution of halo concentrations \citep[\eg][]{Cooray2001, Dolney2004, Giocoli2010}. 

Realistic scatter boosts the matter power at the few per-cent level for $k\gtrsim 10\iMpc$, but may have a larger impact for other spectra or higher-order spectra. It may be that the effect of this can be captured by using the standard deviation of the halo concentration relation, rather than the mean or median. In the spatially-correlated case this scatter is difficult to account for (although the methods of \citealt{Smith2005} could be employed), and falls under the general heading of `assembly bias', requiring its own power spectrum. In contrast, for galaxy power spectra the spatially-uncorrelated scatter in halo-occupation number is usually accounted for (via the variance that emerges from the expectation values in equations~\ref{eq:one_halo_central_central} and \ref{eq:one_halo_satellite_satellite}). If a cross spectrum is calculated there is also a possible covariance between the halo profiles of the two tracers: \cite{Koukoufilippas2020} consider this in the tSZ--galaxy cross correlation as a mass-independent term that then affects the amplitude of the one-halo term. More complicated would be to consider the profile--structure covariance as a function of halo mass.  Finally, we note that for SO haloes there will be no scatter in halo boundary as this is fixed by the halo mass. However, the realistic density field contains triaxial structures and the scatter in the `boundary' of these structures may well contribute additional power in a realistic calculation that accounted for non-spherical haloes.

\section{Modelling beyond-$\Lambda$CDM cosmologies}
\label{sec:altcosmo}

The halo model provides a framework for modelling the matter distribution and its tracers in a way that is not limited to a specific cosmological model. So far, we have assumed that the underlying cosmological model is standard \LCDM. However, the methods discussed can be applied to a variety of alternative cosmologies, as long as we have an estimate of their ingredients: the halo mass function, halo bias, halo profile (and HOD in the case of galaxies). 
The halo model has been used to make predictions for non-linear cosmological observables in a variety of non-standard scenarios. This is achieved either by using physically-motivated arguments to calibrate ingredients that were fitted to \LCDM to the new scenario, or else by running simulations and extracting the required ingredients (hopefully in a form that permits generalisation).  In Section~\ref{sec:ingredients} we saw that the ingredients are usually calibrated against simulations, with the exception of a few cases that use theoretical arguments.  

\subsection{Ingredients measured from simulations}
\label{sec:alt_ingridents}

Simulations have demonstrated that the density profile of dark-matter haloes follow a double power law profile across a wide range of cosmological scenarios.  This is perhaps due to the different modes of mass accretion throughout their life time,  where their initial steady mass accretion phase creates a single power law profile and their secondary violent merger phase flattens their profile in the inner regions \citep[\eg][]{Angulo2017}. Depending on the cosmology, however, different concentration, $c(M)$, relations are observed. These can be extracted from simulations, or somewhat understood via arguments that relate halo concentration to halo formation time \citep[][]{Baldi2010,Diemer2015, Correa2015, Ludlow2016, Diemer2019}. 

We can divide the alternative scenarios into several general categories, ranging from models that modify the properties of dark matter or dark energy, such as warm dark matter (WDM) or dynamical dark energy,  models where gravity is modified,  such as $f(R)$ gravity and scenarios where extra components are included, for example massive neutrinos. 

A number of WDM simulations have been analysed with a focus on measuring halo-model ingredients \citep[for example][]{Schneider2012,Schneider2013,Lovell2014,Ludlow2016}.  WDM particles are expected to wash out structure at smaller scales relative to CDM due to their free streaming, with a characteristic scale dependence linked to the particle mass.  This washing-out effect is seen in the halo-mass function in \nbody simulations. For high-mass haloes, the abundance follows  \LCDM, while there are fewer low-mass haloes.  In addition, while massive haloes follow the same halo profile as in \LCDM, low-mass halos show a relatively steeper profile in their centre (concentration is non-monotonic with mass), suggesting that they have endured fewer mergers.   \cite{Smith2011b} and \cite{Viel2012} consider the halo model in WDM models,  and find that predictions are improved if a genuinely smooth component of the matter is assumed and accounted for in the bias and mass function (see Subsection~\ref{sec:smooth}).  There are indications of an enhanced linear halo bias from both simulations and theoretical work.  Adding all of these new components into a halo model approach, \cite{Schneider2012} find a 10\% agreement with the measured power spectra from simulations.  This may be further improved with the addition of non-linear halo bias (see Section~\ref{sec:non_linear_bias}).  Aside from WDM, other dark matter models have been explored in the literature, for example \cite{Simon2021, Dome2023} simulated fuzzy dark matter.  

Dynamical dark energy is perhaps the most well-known extension to \LCDM. As a result many simulations include one or two equation-of-state parameters that describe the density variation of dark energy with time \citep[for example][]{Heitmann2016} and can capture the behaviour of a wide range of such models, where dark energy is minimally coupled  \cite[see][for quintessence with Ratra-Peebles and SUGRA potentials]{DEUSS}. These models generally only affect the background cosmology,  but their impact can be seen in the growth of structures.  A dark energy with a negative equation-of-state parameter,  acts as an anti-gravity force emptying voids and creating a higher contrast cosmic web.  Simulations find a relative universality relation for the halo mass function, at the  $10\%$ level,  once the linear growth function is adjusted to include the impact of these dark energy models \citep{Courtin2011,Heitmann2016}. Other dark-energy models, such as clustering dark energy have also been explored \citep[see ][which focuses on cluster counts]{Batista2017}.

Dark energy may also interact with dark matter through an energy exchange which couples these fields.  The energy can flow either direction,  but most interesting are models were dark matter is transformed into dark energy,  as they can potentially explain the dominance of dark energy in the current epoch.  \cite{Baldi2010,Baldi2023} simulations show that, as expected, the energy transfer from dark matter to dark energy reduces the abundance of haloes as well as halo concentration, across all masses.   Such models have also been explored via spherical collapse in the literature,  finding qualitatively consistent results \citep{Wintergerst2010,Tarrant2012, Barros2019,Barros2020}.

Alternative gravity models that were originally introduced to account for the origins of $\Lambda$, have more recently been explored to test the validity of general relativity at cosmological scales.  These models require a screening mechanism so that they reduce to the general relativity solutions at smaller scale and/or denser regions in the Universe.  Simulations for the more popular models exist today.  For example $f(R)$ has been simulated in \cite{Schmidt2009a,Li2012},  showing that the abundance of rare massive halos increases while their bias is decreased and their profile remains unchanged (this is preserved by the screening mechanism),  as a result power spectra are enhanced in a scale dependant manner. \citeauthor*{Dvali2000} (DGP; \citeyear{Dvali2000}) gravity was explored in \cite{Schmidt2010a, Schmidt2010b} who find qualitative agreement between spherical collapse and simulations.  \cite{Barreira2014} simulate two types of Galileon gravity and explored their impact on halo model ingredients.

Massive neutrinos are a form of (subdominant) hot dark matter that can wash out structures on larger scales compared to WDM.  The impact of massive neutrinos on structures has been studied in simulations assuming varying levels of model complexity, starting from background only effects to including neutrino particles and evolving them alongside the cold dark matter particles \citep{Brandbyge2008, Brandbyge2009,Viel2010, Agarwal2011, Bird2012,  Massara2014,  Upadhye2014}.  Since modelling the behaviour of the massive neutrino particles is challenging, these studies disagree on the best methodology and their associated accuracy. 

The impact of these models can be included in the halo model to reach percent-level agreement for power spectra when compared to simulation results.  \cite{Mead2016}, for example, models the effect of several alternative cosmologies within the \hmcode formalism: dynamical dark energy (calibrated against their own simulations),   coupled dark matter-dark energy (calibrated against \citealt{Baldi2010}),  massive neutrinos (calibrated against \citealt{Massara2014}), $f(R)$ gravity (calibrated against \citealt{Li2013}).  They find a few percent agreement in all cases, aside from the \cite{Vainshtein1972} screening scenario, which achieves a $10\%$ accuracy.   \cite{Dentler2022} also used the \hmcode formalism to set constraints on fuzzy dark matter from data.  Note that \hmcode only provides matter power, and cannot easily be generalised to other tracers. 

Most simulations of alternative models do not include baryonic effects such as AGN feedback. There are, however, a limited number of studies that have explored hydrodynamical simulations for alternative models.   \cite{Puchwein2013b} introduce a code for hydrodynamical simulation of modified gravity models and show some degeneracies on the power spectrum between the effects of baryonic feedback and $f(R)$ gravity.  The impact of $f(R)$ gravity on abundance and other properties of clusters is further investigated by \cite{Arnold2014}. Other authors have included baryons as a separate component in their simulations to account for the differing coupling mechanisms between baryons and other particles, although they do not account for the full hydrodynamics of baryons \citep[for example][]{Baldi2010,Hammami2015,Baldi2022}.

\subsection{Spherical collapse}
\label{sec:spherical_collapse}

The most accurate understanding of how structure formation proceeds in non-standard cosmologies derives from simulations. However, considering the collapse of a single spherical perturbation in an otherwise featureless universe, the so-called `spherical-collapse model', has provided valuable insights about the formation process, effective linear-collapse threshold, and the virial radius of collapsed haloes.

The differential equation governing the collapse of a perturbation of uniform overdensity $\delta$ is
\begin{equation}
\ddot\delta+2H\dot\delta-\frac{4}{3}\frac{\dot\delta^2}{1+\delta}=\frac{3}{2}\Omega_\mathrm{m}(a)H^2\delta(1+\delta)\ ,
\label{eq:spherical_collapse}
\end{equation}
where dots denote time derivatives and it has been assumed that an initially uniform perturbation remains uniform throughout its evolution. $H$ is the (time-dependent) Hubble parameter and $\Omega_\mathrm{m}(a)$ is the (time-dependent) value of $\Omega_\mathrm{m}$. Usually (but not always\footnote{Some perturbations can be prevented from collapsing as dark energy, or similar, comes to dominate the expansion.}) solving equation~(\ref{eq:spherical_collapse}) for an initially small perturbation reveals that the perturbation grows, reaches a maximum size, and then collapses (due to the spherical symmetry this collapse is to an infinite-density singularity). Equation~(\ref{eq:spherical_collapse}) can be linearised to give
\begin{equation}
\ddot\delta+2H\dot\delta=\frac{3}{2}\Omega_\mathrm{m}(a)H^2\delta\ ,
\label{eq:linear_theory}
\end{equation}
which is the familiar differential equation governing the evolution of matter perturbations on sub-horizon scales. 

Equations~(\ref{eq:spherical_collapse}) and (\ref{eq:linear_theory}) can be solved in tandem, for the same initial condition, to reveal what value linear theory would predict when the non-linear theory reaches collapse. In an Einsten--de Sitter universe this value is the oft-quoted $\delta_\mathrm{c}\simeq1.686$, but this value has some cosmology dependence, for example in an $\Omega_\mathrm{m}=0.3$ \LCDM universe $\delta_\mathrm{c}\simeq1.676$, lower by $\sim 0.5\%$ \citep[\eg][]{Nakamura1997}.  As we discussed at the end of Section~\ref{sec:halo_mass_function},  including the cosmology dependence of $\delta_\mathrm{c}$ can improve the mass-function universality and halo-model predictions. 

Of course, real perturbations are not perfectly spherical, and even if they were, inhomogeneity in the true gravitational field would distort them from this idealised form. No perturbation will collapse to a singularity but instead will equilibrate into a roughly spherical structure with a density gradient and with kinetic and potential energy split as per the virial theorem (a virialized halo). If one takes the point of halo formation as given by the time the spherical-model would predict a singularity to form, one can then derive a rough estimate for the density of the resulting structure by applying the virial theorem at the point of collapse to derive a radius. In an Einstein--de Sitter universe this gives the result that the halo overdensity should be $\Delta_\mathrm{v}\simeq178$, independent of halo mass. This is the origin of the $\sim 200$ overdensity criterion that is often used when identifying haloes in simulations. In an $\Omega_\mathrm{m}=0.3$ model $\Delta_\mathrm{v}\simeq 330$ or $\simeq310$ depending on whether the contribution of dark energy is included within the virial theorem or not.  Fitting functions for the cosmology dependence of $\Delta_\mathrm{v}$ can be found in \citeauthor{Bryan1998} (\citeyear{Bryan1998}; \LCDM) and \citeauthor{Mead2017} (\citeyear{Mead2017}; homogeneous dark energy, ignoring dark-energy contributions to virialization).

Analogues to equation~(\ref{eq:spherical_collapse}) can also be derived and solved in any fully-specified cosmological model.  This is done for dark energy models \citep{Mota2004, Bartelmann2006, Abramo2007, Pace2010, Wintergerst2010, Mead2017, Barros2020},  but care must be taken with dark energy perturbations which can be important if its sound speed is smaller than speed of light \citep{Batista2021,Batista2023}. In early dark energy scenarios the initial conditions for the equations need to be set in the correct linear growing mode. In coupled dark energy--dark matter models the coupling needs to be incorporated, as well as the back reaction of this on the local dark energy. Spherical collapse in massive-neutrino cosmologies was investigated by \cite{LoVerde2014}. In viable modified gravity models \citep[\eg][]{Brax2010, Schmidt2010a, Schmidt2010b, Borisov2012, Lombriser2013a, Kopp2013, Barreira2013, Taddei2014} screening mechanisms must be incorporated. Spherical collapse in generalised dark matter has been investigated by \cite{Pace2020}.

\subsection{\react}
\label{sec:react}

While the raw halo model predictions for the matter spectrum have been shown to be inaccurate at the $\sim 10\%-30\%$ level \citep[\eg][]{Mead2015b}, it has been demonstrated that the ratio of halo model predictions for different cosmologies can more accurately predict the same ratio measured in simulations \citep[\eg][]{Schmidt2010a, Schmidt2010b, Mead2017,Gupta2023}. This is particularly true if the two cosmologies are chosen such that they share a linear spectrum in both shape and amplitude, and if the spherical-collapse model is used to include the cosmology dependence of $\delta_\mathrm{c}$ and $\Delta_\mathrm{v}$. 

Originally presented by \cite{Cataneo2019}, the \react method uses this insight to produce accurate spectra for dynamical dark energy and modified gravity models ($f(R)$ and DGP), but also incorporates perturbation theory.  Within the \react formalism a pseudo power spectrum needs to be defined, which has the same shape and amplitude as the linear power spectrum in the target cosmology,
\begin{equation}
P^{\rm psuedo}_{\rm linear}(k) = P^{\rm target}_{\rm linear}(k)\;.
\end{equation}
Although there may not be a match for the pseudo linear power spectrum under the flat-\LCDM model,  its evolution to non-linear power spectrum assumes a flat-\LCDM model without neutrinos.  This evolution can be calculated using \hmcode or emulators \citep[\eg][]{Giblin2019}.  The non-linear power spectrum is then constructed via
\begin{equation}
P^{\rm target}_{\rm non-linear}= R(k) P^{\rm psuedo}_{\rm non-linear}(k)\;,
\end{equation}
where $R(k)$ is the `reaction'.  To predict the reaction a combined halo model and perturbation theory approach is employed \citep[see][for a concise review]{Cataneo2022}.  Note that \react currently only targets the matter spectra and only predicts the reaction of the power spectrum to a particular ingredient, and it therefore requires accurate non-linear spectra for \LCDM as an ingredient.

Updates to \react have been presented by \citeauthor{Cataneo2020} (\citeyear{Cataneo2020}; massive neutrinos), \citeauthor{Bose2020} (\citeyear{Bose2020}; modified gravity), and \citeauthor{Bose2021} (\citeyear{Bose2021}; massive neutrinos and baryonic feedback).  \cite{Srinivasan2021} have applied the reaction methodology to time dependent $\mu$,  which captures the phenomenology of a range of modified gravity models. The \react formalism was applied to cosmological analysis of the KiDS weak lensing data to set constraints on $f(R)$ gravity \citep{Troester2021}.

\section{Software}
\label{sec:software}

Together with this paper, we provide a \textsc{python} package that can be used to implement the calculations described here: \pyhalomodel \footnote{\link{https://github.com/alexander-mead/pyhalomodel}}. Using the same software we have also written a pure-\python implementation of \href{https://github.com/alexander-mead/HMcode-python}{\sc hmcode}\footnote{\link{https://github.com/alexander-mead/HMcode-python}}, although this has not been used to make figures for this paper. There are several other software packages that are publicly available and that we have investigated while writing this review: 
\begin{itemize}

\item The \textsc{dark emulator}\footnote{\link{https://darkquestcosmology.github.io/}} of \cite{darkemu} emulates the halo mass function and power spectrum, and has the capacity to perform galaxy--galaxy and galaxy--matter halo-model calculations on top of the emulated quantities. Uniquely, because the halo power spectrum is emulated, the non-linear halo bias (called $\Bnl$ in this paper) is automatically included.

\item Recently, the \halomod\footnote{\link{https://github.com/halomod/halomod}} package of \cite{halomod} has been released, which can be used as either a \python package or as a browser application\footnote{\link{https://thehalomod.app}}. Currently, this can be used for matter and galaxy power spectra, but the code is extendable to other tracers in principle.

\item \textsc{colossus}\footnote{\link{https://bdiemer.bitbucket.io/colossus/index.html}} \citep{Diemer2018} is a \python toolkit for calculations pertaining to cosmology, the large-scale structure of the universe, and the properties of dark matter haloes. It does not perform halo-model calculations, but does provide the necessary ingredients from a variety of different sources.

\item The LSST Core Cosmology Library \citep[\textsc{ccl}\footnote{\link{https://github.com/LSSTDESC/CCL}};][]{CCL} contains a standard halo model calculator written in \python.

\item The \hmcode\footnote{\link{https://github.com/alexander-mead/HMcode}} software, developed by \citep{HMcode} performs a version of the standard halo model calculation for the matter power spectrum. The calculation is augmented for enhanced accuracy. The original source code is written in \textsc{fortran}, but a \python wrapper\footnote{\link{https://pypi.org/project/pyhmcode/}} around the \textsc{fortran} is also available. \hmcode is also included within \camb\footnote{\link{camb.readthedocs.io}}.

\end{itemize}

Other public codes that we are aware of, but that we have not had a chance to thoroughly investigate are \textsc{halogen}\footnote{\link{https://github.com/EmmanuelSchaan/HaloGen}} \citep[][]{halogen},  \textsc{aum}\footnote{\link{https://github.com/surhudm/aum}} \citep{More2015a} and \textsc{class\textunderscore sz} \footnote{\link{https://github.com/CLASS-SZ/class\textunderscore sz}}  \citep{Bolliet2018, Bolliet2023}.

\section{Summary}
\label{sec:summary}

We have presented a pedagogical review of the halo model and of its uses in analysing cosmological data.  The flexibility of this framework, and its ability to model non-linear scales, makes it very attractive for analysing data from a multitude of probes of large-scale cosmological structure, and to extract information from smaller scales where linear perturbation theory no longer applies.  


We began with a derivation of the power spectrum and introduced the concept of two- and one-halo terms.  We continued by describing how the power spectrum is modified for discrete tracers of matter,  such as galaxies (Section~\ref{sec:basics}).  To make a prediction with the halo model one needs estimates of its ingredients: the abundance of haloes of different masses (halo mass function,  HMF),  the relation between the halo distribution and the underlying linear matter field (halo bias) and the distribution of matter or its tracers within the haloes (halo profile).  Aside from a few exceptions, these ingredients are measured and calibrated against simulations.  In Section~\ref{sec:ingredients} we focused on modelling matter power spectra and the ingredients needed to do so,  including the modelling of baryon feedback and versions of the halo model that are directly calibrated against simulations,  such as \halofit and \hmcode.    

We then turned our focus to the modelling of tracers of matter in Section~\ref{sec:tracers}.  Once we know how a tracer populates the haloes we can predict its distribution using the same halo model formalism.  We began with galaxies and halo occupation distribution (HOD) and how central and satellite galaxies need to be treated separately.  Aside from galaxy positions,  large-scale structure impacts the intrinsic alignments of galaxies which can be modelled in a similar manner.  While galaxies are discrete tracers for matter,  other tracers, such as hot electrons in galaxy clusters, seen for example through the tSZ effect,  are diffused tracers.  We further discussed a few examples of such diffuse signals. 

The standard `vanilla' halo model makes simplifying assumptions,  such as that all haloes are spherical and that they are linearly biased with respect to matter.  In Section~\ref{sec:non_standard} we relax some of these assumptions and discuss what improvements are expected when adding more complexity to the base halo model formalism.  We note that some of these improvements are more important for two-point functions of specific tracer combinations. 

While the majority of the review is focused on solutions within the standard \LCDM cosmologies,  in Section~\ref{sec:altcosmo} we show that the same approach can be applied to exotic cosmological models and extensions to \LCDM; for example,  models with massive neutrinos,  $f(R)$ gravity and dynamical or interacting dark energy models.  In these cases the spherical-collapse model can provide useful insights.  In particular,  we discuss that although the halo model can be inaccurate in predicting power spectra,  it can estimate the ratios of non-linear power spectra at percent level,  when the linear power spectra are matched between the two models.  The spherical-collapse solutions can then be applied to include the cosmology dependence of key inputs for the halo model.  The \react formalism, which we briefly introduce at the end of Section~\ref{sec:altcosmo}, takes full advantage of this property of the halo model. 

Finally, we release a halo model code including \textsc{jupyter} notebooks for (almost) all figures in this paper and many demo examples: \pyhalomodel. In Section~\ref{sec:software} we also listed other halo model related codes that we are aware of.

The use of the halo model for cosmological analysis has seen a recent boost \citep[for example][]{Miyatake2022b,Troester2022, Dvornik2023,Amon2023},  thanks to creative applications of corrections and adjustments to this flexible framework.  The halo model can be applied in a more analytic versus a more simulation based approach.  While simulation-based approaches can be more accurate, they are limited to the range of models that are simulated, and therefore lose some of the flexibility of the halo model.  Many recent analyses, however, opt for a half-way method that inherits the best of both worlds.  New methods of applying the halo model, for example \react or the inclusion of non-linear halo bias have the potential to address  most or all of the inaccuracies in the vanilla halo model predictions, while maintaining flexibility.  With these advancements,  we expect the halo model to be applied with more rigour to upcoming cosmological survey data,  which was the main motivation for writing this review.  Finally, we note that under the halo-model approach there is a pathway for uniting astrophysics and cosmology.

\section*{Acknowledgements}

We are thankful for comments from Shahab Joudaki and Constance Mahony.  We are grateful for the constructive input from the community and an anonymous referee.  We acknowledge support from the European Research Council under grant number 647112 (MA,AM,CH), from the Max Planck Society and the Alexander von Humboldt Foundation in the framework of the Max Planck-Humboldt Research Award endowed by the Federal Ministry of Education and Research (AM, CH), and the UK Science and Technology Facilities Council (STFC) under grant ST/V000594/1 (MA,CH).

\bibliographystyle{mnras} 
\bibliography{biblio}

\end{document}